%% file: SMP-20-004_temp.tex
\pdfoutput=1
\documentclass[11pt,twoside,a4paper,cmspaper,final,collab]{cms-tdr}
\def\svnVersion{9f0f381}\def\svnDate{2025/04/29}\def\cmsCernNoTag{CERN-EP-2024-134}\def\cmsCernDate{\today}\def\cmsMessage{Published in the Journal of High Energy Physics as \href{http://dx.doi.org/10.1007/JHEP04(2025)162}{\doi{10.1007/JHEP04(2025)162}.}}
\begin{document}\cmsNoteHeader{SMP-20-004}

\newcommand{\pb}{\ensuremath{\,\text{pb}}\xspace}

\newlength\cmsTabSkip\setlength{\cmsTabSkip}{1ex}
\providecommand{\cmsTable}[1]{\resizebox{\textwidth}{!}{#1}}

\cmsNoteHeader{SMP-20-004}

\title{Measurement of the inclusive cross sections for \PW and \PZ boson production in proton-proton collisions at \texorpdfstring{$\sqrt{s}=5.02$}{sqrt(s) = 5.02} and \texorpdfstring{13\TeV}{13 TeV}}

\date{\today}

\abstract{Measurements of fiducial and total inclusive cross sections for \PW and \PZ boson production are presented in proton-proton collisions at $\sqrt{s}=5.02$ and 13\TeV. Electron and muon decay modes ($\Pell=\Pe$ or \PGm) are studied in the data collected with the CMS detector in 2017, in dedicated runs with reduced instantaneous luminosity. The data sets correspond to integrated luminosities of $298\pm6\pbinv$ at 5.02\TeV and $206\pm5\pbinv$ at 13\TeV. Measured values of the products of the total inclusive cross sections and the branching fractions at 5.02\TeV are $\sigma(\Pp\Pp \to \PW + \PX)\mathcal{B}(\PW \to \Pell\PGn) = 7300 \pm 10 \stat \pm 60 \syst \pm 140 \lum \pb$, and $\sigma(\Pp\Pp \to \PZ + \PX)\mathcal{B}(\PZ\to\Pell^+\Pell^-) = 669 \pm 2 \stat \pm 6 \syst \pm 13 \lum\pb$ for the dilepton invariant mass in the range of 60--120\GeV. The corresponding results at 13\TeV are $20480\pm 10\stat\pm 170\syst\pm 470\lum\pb$ and $1952\pm 4\stat \pm 18\syst\pm 45\lum\pb$. The measured values agree with cross section calculations at next-to-next-to-leading-order in perturbative quantum chromodynamics. Fiducial and total inclusive cross sections, ratios of cross sections of $\PWp$ and $\PWm$ production as well as inclusive \PW and \PZ boson production, and ratios of these measurements at 5.02 and 13\TeV are reported.}

\hypersetup{%
    pdfauthor={CMS Collaboration},%
    pdftitle={Measurement of the inclusive cross sections for W and Z boson production in proton-proton collisions at sqrt(s) = 5.02 and 13 TeV},%
    pdfsubject={CMS},%
    pdfkeywords={CMS, W/Z bosons}
}

\maketitle

\section{Introduction}
Measurements of the inclusive cross sections for \PW and \PZ boson production at hadron colliders are important tests of higher-order perturbative quantum chromodynamics (QCD) and electroweak (EW) corrections. The precise determination of the inclusive cross sections and their ratios provides constraints on the parton distribution functions (PDFs) of the proton~\cite{Butterworth:2015oua,Bailey:2020ooq,Hou:2019qau,NNPDF:2017mvq}. Theoretical predictions of the inclusive cross sections for \PW and \PZ boson production are available at next-to-next-to-leading order (NNLO)~\cite{Melnikov:2006kv,Catani:2009sm} and next-to-NNLO (N$^{3}$LO)~\cite{Baglio:2022wzu} accuracy in QCD. The EW corrections are available at next-to-leading-order (NLO) accuracy~\cite{Anastasiou:2003ds, Dittmaier:2014qza, Lindert:2017olm}.

This paper describes the measurements of the products of the inclusive cross sections for \PW and \PZ boson production and leptonic ($\Pell=\Pe$ or \PGm) decay branching fractions at the CERN LHC with the CMS detector~\cite{Chatrchyan:2008zzk}. The gauge bosons are reconstructed via their decays to electrons and muons, which provide a clean experimental signature. The $\PZ/\PGg^{*} \to \Pell^+\Pell^-$ process is referred to as the \PZ boson process in this paper.

The measurements are performed in proton-proton ($\Pp\Pp$) collisions at $\sqrt{s}=5.02$ and 13\TeV, collected at the end of 2017 in dedicated runs with reduced instantaneous luminosity. The peak delivered instantaneous luminosities were $1.3\times10^{33}\unit{cm}^{-2}\unit{s}^{-1}$ and $2.5\times10^{33}\unit{cm}^{-2}\unit{s}^{-1}$ with an average of two and three $\Pp\Pp$ interactions per bunch crossing at $\sqrt{s}=5.02$ and 13\TeV. Because of this, the missing transverse momentum (\ptmiss) resolution and the lepton isolation efficiency in these special runs are improved compared with those in regular runs, therefore reducing the background contamination after final event selections. The total integrated luminosities of these special runs used in the analysis described in this paper are 298\pbinv at 5.02\TeV and 206\pbinv at 13\TeV. The measurements of cross section ratios of \PWp and \PWm production as well as inclusive \PW and \PZ boson production, and ratios of these measurements at 5.02 and 13\TeV are also performed.

Fiducial and total inclusive cross sections for \PW and \PZ boson production and cross section ratios were previously studied by the ATLAS, CMS, and LHCb Collaborations in $\Pp\Pp$ collisions at $\sqrt{s}=2.76$, 5.02, 7, 8, 13, and 13.6\TeV at the LHC~\cite{Chatrchyan_2015,CMS:2012fgk,Aad:2019bdc,Aaboud:2018nic,Aaboud:2016btc,CMS:2011aa,Aaij:2014wba,Aaij:2012mda,Aaij:2015gna,Aaij:2016qqz,Aaij:2016mgv,Aaij:2015zlq,Chatrchyan:2014mua,Aad:2016naf,Aaij:2016mgv,CMS:2019raw,CMS:2020cph,ATLAS:2024nrd,ATLAS:2024irg}. Compared with the previous results, the current measurements have reduced systematic uncertainties, and include the ratios of the production cross sections measured at two center-of-mass energies (5.02\TeV and 13\TeV). Tabulated results are provided in the HEPData record for this analysis~\cite{hepdata}.

\section{The CMS detector}
\label{sec:CMSdetector}
The central feature of the CMS apparatus is a superconducting solenoid
of 6\unit{m} internal diameter, providing a magnetic field of
3.8\unit{T}. Within the magnetic volume there are a silicon pixel and strip
tracker, a lead tungstate crystal electromagnetic calorimeter (ECAL),
and a brass and scintillator hadron calorimeter, each composed
of a barrel and two endcap sections. Forward calorimeters extend the pseudorapidity ($\eta$)
coverage provided by the barrel and endcap detectors. Muons are detected in gas-ionization
detectors embedded in the steel flux-return yoke outside the solenoid. A more detailed
description of the CMS detector, together with a definition of the coordinate system used
and the relevant kinematic variables, is presented in Ref.~\cite{Chatrchyan:2008zzk}.

Events of interest are selected using a two-tiered trigger system. The first level, composed of custom hardware processors, uses information from the calorimeters and muon detectors to select events of interest at a rate of around 100\unit{kHz} within a fixed latency of 4\mus~\cite{CMS:2020cmk}. The second level, known as the high-level trigger,
consists of a farm of processors running a version of the full event reconstruction
software optimized for fast processing, and reduces the event rate to $\mathcal{O}$(1\unit{kHz})
before data storage~\cite{Khachatryan:2016bia}.

\section{Simulated event samples}
\label{sec:Samples}
Several Monte Carlo (MC) event generators are used to simulate the signal and background
processes. The \PW and \PZ boson signal samples ($\PW\to\Pell\PGn$ and $\PZ\to\Pell^{+}\Pell^{-}$, where $\Pell=\Pe, \PGm$) at 5.02 and 13\TeV are simulated at NLO in perturbative QCD with the \MGvATNLO event generator in version 2.3.3~\cite{Alwall:2014hca}. The same settings are used for the simulation of $\PW\to\PGt\PGn$ and $\PZ\to\PGt^{+}\PGt^{-}$ background processes.
Other background samples are simulated with \POWHEG~2.0~\cite{Nason:2004rx,Frixione:2007vw,Alioli:2008gx,Alioli:2010xd}, including the diboson ($\PW\PW$, $\PZ\PZ$, and $\PW\PZ$) and $\ttbar$ processes~\cite{Chiesa:2020ttl,Frixione:2007nu}. The \PYTHIA~8.230~\cite{Sjostrand:2014zea} package is used for parton showering, hadronization, and the underlying event simulation, with the CP5 tune~\cite{Skands:2014pea, Sirunyan:2019dfx}. The NNPDF~3.1~\cite{NNPDF:2017mvq} set of PDFs at NNLO in QCD is used as the default set of PDFs.

The detector response is simulated using a detailed description of the CMS detector based on the \GEANTfour package~\cite{Agostinelli:2002hh}, and event reconstruction is performed with the same algorithms used for the observed events. Additional $\Pp\Pp$ interactions (pileup) occurring in the same or nearby beam crossing as the event of interest are included in the simulation.

\section{Event reconstruction}
\label{sec:reconstruction}
The particle-flow (PF) algorithm~\cite{Sirunyan:2017ulk} reconstructs and identifies each individual particle in an event, with an optimized combination of all subdetector information. The individual particles (PF candidates) are identified as charged and neutral hadrons, leptons, or photons.
The primary vertex is taken to be the vertex corresponding to the hardest scattering in the event, evaluated using tracking information alone, as described in Section 9.4.1 of Ref.~\cite{CMS-TDR-15-02}.

Hadronic jets are clustered from these reconstructed particles using the infrared and collinear safe anti-\kt algorithm~\cite{Cacciari:2008gp, Cacciari:2011ma} with a distance parameter of 0.4.
The vector \ptvecmiss is computed as the negative vector sum of the transverse momenta of all the PF candidates in an event. Its magnitude is denoted as \ptmiss~\cite{CMS:2019ctu}.

Electrons are reconstructed by associating a track reconstructed in the inner silicon detectors with a cluster of energy in the ECAL~\cite{CMS:2020uim,CMS-DP-2020-021}. The electron momentum is estimated by combining the energy measurement in the ECAL with the momentum measurement in the tracker. The momentum resolution for electrons with $\pt \approx 45\GeV$ from $\PZ \to \Pe \Pe$ decays ranges from 1.6 to 5\%. It is generally better in the barrel region than in the endcaps, and also depends on the bremsstrahlung energy emitted by the electron as it traverses the material in front of the ECAL. The selected electron candidates cannot originate from photon conversions in the inner silicon tracker material and must satisfy a set of quality requirements based on the shower shape of the energy deposit in the ECAL. The ECAL barrel-endcap transition region $1.44 < \abs{\eta} < 1.57$ is excluded to avoid the region with poor electron reconstruction efficiency and resolution.

Muons are reconstructed by associating a track reconstructed in the inner silicon detectors with a track in the muon system. Muons are measured in the range $\abs{\eta} < 2.4$, with detection planes made using three technologies: drift tubes, cathode strip chambers, and resistive plate chambers. The single muon trigger efficiency exceeds 90\% over the full $\eta$ range, and the efficiency to reconstruct and identify muons is greater than 96\%. Matching muons to tracks measured in the silicon tracker results in a relative transverse momentum resolution, for muons with \pt up to 100\GeV, of 1\% in the barrel and 3\% in the endcaps. The \pt resolution in the barrel is better than 7\% for muons with \pt up to 1\TeV~\cite{Sirunyan:2018fpa}. The selected muon candidates are required to satisfy a set of quality requirements based on the number of spatial measurements in the silicon tracker and the muon system, as well as the fit quality of the combined muon track~\cite{Sirunyan:2018fpa}.

The lepton (electron or muon) candidate tracks must be consistent with the primary vertex of the event to suppress electron candidates from photon conversions and lepton candidates originating from decays of heavy quarks and $\tau$ leptons. The lepton candidates must be isolated from other particles in the event. The relative isolation for the lepton candidates with $\pt^{\Pell}$ is defined as
\begin{equation}
        R_\text{iso} = \bigg[\sum_{\substack{\text{charged} \\ \text{hadrons}}} \!\! \pt \, + \,
            \max\big(0, \sum_{\substack{\text{neutral} \\ \text{hadrons}}} \!\! \pt
            \, + \, \sum_{\text{photons}} \!\! \pt \, - \, \pt^\mathrm{PU}
            \big)\bigg] \bigg/ \pt^{\Pell},
        \label{eq:iso}
\end{equation}
where the sums run over the charged hadrons, neutral hadrons, and photons in a cone defined by $\Delta R \equiv \sqrt{\smash[b]{(\Delta\eta)^2 + (\Delta\phi)^2}} = 0.3$ (0.4) around the electron (muon) trajectory. To mitigate the contribution of pileup interactions to the isolation, only charged hadrons originating from the primary vertex are included in the first sum~\cite{CMS:2020ebo}, and the estimated neutral contribution from pileup, $\pt^\mathrm{PU}$, is subtracted~\cite{CMS:2020uim,Sirunyan:2018fpa} from the isolation quantities. For electrons, $\pt^\mathrm{PU}$ is estimated as $\rho A_\mathrm{eff}$, where $\rho$ is the median of the transverse energy density per unit area in the event and $A_\mathrm{eff}$ is the area of the isolation region weighted by a factor that accounts for the dependence of the pileup transverse energy density on the object $\eta$~\cite{CMS:2020uim}. For muons, $\pt^\mathrm{PU}$ is estimated by $0.5\sum_i \pt^{\mathrm{PU},i}$, where $i$ runs over the charged hadrons originating from pileup vertices, and the factor 0.5 corrects for the ratio of charged to neutral particles in the isolation cone~\cite{Sirunyan:2018fpa}.

\section{Event selection}
\label{sec:selection}
Leptonic \PW boson decays are characterized by a prompt, energetic, and
isolated charged lepton and a neutrino giving rise to significant \ptmiss. The \PZ boson decays to leptons are selected by requiring two energetic and isolated leptons of the same flavor and opposite charge.

The events are triggered by the presence of at least one electron with transverse momentum $\pt>17 (20)\GeV$ and $|\eta|<2.5$ at 5.02 (13)\TeV, or at least one muon with $\pt>17\GeV$ and $\abs{\eta}<2.4$. The triggered electron and muon candidates satisfy less restrictive isolation and quality requirements than the offline selection criteria.

The fiducial region for the \PW boson is defined by the charged-lepton kinematic requirements with $\pt > 25\GeV$ and $\abs{\eta} < 2.4$ and a requirement on the \PW boson transverse mass $m_{\mathrm{T}}>40\GeV$, defined as $m_{\mathrm{T}}=\sqrt{\smash[b]{2\pt^{\Pell}\pt^{\PGn}[1-\cos(\Delta\phi)]}}$, where $\pt^{\PGn}$ is the magnitude of the neutrino \pt and $\Delta\phi$ is the azimuthal angle between the lepton and the neutrino directions. The $\pt^\PGn$ is estimated by $\ptmiss$. To reduce contamination from events with a \PZ boson, events are rejected if there are additional muons or electrons with $\pt>10\GeV$ that satisfy less restrictive selection requirements (with average selection efficiencies above 95\%) than the signal lepton candidate selection~\cite{Sirunyan:2018fpa,CMS:2020uim}.

The fiducial region for the \PZ boson is defined by a common set of kinematic selections applied to both the $\Pep\Pem$ and $\PGmp\PGmm$ final states at generator level, emulating the selection performed at the reconstruction level. Leptons are required to have $\pt > 25\GeV$ and $\abs{\eta} < 2.4$, and the dilepton invariant mass ($m_{\Pell^{+}\Pell^{-}}$) is required to lie within $60$-$120\GeV$.

The products of the fiducial inclusive production cross sections for the \PW and \PZ bosons and their leptonic decay branching fractions can be expressed as:
\begin{equation}
    \sigma^{\text{fid}}\mathcal{B} = \frac{N_{\mathrm{sig}}}{\epsilon\lumi}
     \label{eq:xsec}
\end{equation}
where $N_{\mathrm{sig}}$ is the number of signal events, $\epsilon$ is the efficiency of the signal events within the fiducial region satisfying the event selection requirements, and \lumi is the integrated luminosity. The product of the total inclusive cross section $\sigma^{\mathrm{tot}}$ and $\mathcal{B}$ is defined as $\sigma^{\text{fid}}\mathcal{B}/A$, where $A$ is the kinematic acceptance. Electrons and muons produced in the decay of $\PGt$ leptons are excluded in the definition of the fiducial region.
The leptons at the generator level are defined at Born level, \ie, before the quantum electrodynamics final-state radiation (FSR).

The efficiencies for the reconstruction, identification, and isolation
requirements on the leptons are obtained in bins of charge, \pt, and $\eta$ using the
``tag-and-probe'' technique~\cite{CMS:2011aa} with $\PZ\to\Pell^{+}\Pell^{-}$ events that provide an unbiased data sample with high purity. The electron (muon) candidates have an average combined reconstruction, identification, and isolation efficiency of 70 (90)\%. Electron momentum scale corrections~\cite{CMS:2020uim} depend on the number of simultaneous collisions, and are therefore derived using special low pileup runs at 13\TeV. The dependence on the center-of-mass energy is small and therefore the same electron momentum scale corrections are applied at 5.02\TeV and 13\TeV. Muon momentum scale corrections~\cite{Bodek:2012id} are derived from regular runs in 2017 and are applicable to special runs in this inclusive cross section analysis.

During the 2017 data taking, a gradual shift in the timing of the inputs of the ECAL first-level trigger in the region at $\abs{\eta} > 2.0$ caused a specific trigger inefficiency, referred to as ``prefiring"~\cite{CMS:2024ppo}. For events containing an electron (a jet) with \pt larger than $\approx$50\GeV ($\approx$100\GeV), in the region $2.5 < \abs{\eta} < 3.0$, the efficiency loss is $\approx$10--20\%, depending on \pt, $\eta$, and time. Correction factors were computed from data and applied to the acceptance evaluated by simulation.

\section{Signal extraction and background estimation}
\label{sec:backgrounds}
The statistical analysis of the event yields is performed with a binned maximum likelihood fit to the $m_{\mathrm{T}}$ and $m_{\Pell^{+}\Pell^{-}}$ distributions in the \PW and \PZ boson signal regions, respectively. The fits are done simultaneously in the electron and muon channels, assuming lepton flavor universality. Electron and muon channels are also fitted separately and the results are compatible within uncertainty. The systematic uncertainties are treated as nuisance parameters in the fit. A combination of methods based on control samples in data and simulation studies is used to estimate background contributions. In all cases where simulated samples are used, events are reweighted to correct the lepton and trigger efficiencies to match those measured in data.

The Drell--Yan (DY) lepton pair production is a background in the \PW boson signal selection when one of the two leptons does not enter the fiducial region or is not reconstructed. The contribution of this background process as well as the contribution of the top quark and diboson processes are estimated from the simulation, and normalized to their expected cross sections with the corresponding theoretical and experimental uncertainties. The contribution of the signal events not entering the fiducial region at the generator level is normalized to the corresponding signal yield in the final likelihood fit through the ratios of the theoretical cross sections. The same treatment is applied to estimate the contributions from the $\PW\to\PGt\PGn$ and $\PZ\to\PGt^{+}\PGt^{-}$ backgrounds. The background contribution from $\PW \to \PGt\PGn$, DY, and diboson processes is denoted as the EW background.

Another source of background in the \PW boson signal region is the standard model events composed uniquely of jets produced through the strong interaction, referred to as QCD multijet events. An accurate simulation of the \ptmiss is essential to distinguish the \PW boson signal from the QCD multijet backgrounds. The \ptmiss response and resolution is derived in the \PZ boson signal region. The hadronic recoil, which is the hadronic system the boson recoils against, is studied in data and simulations after the \PZ boson kinematic selections, and then the measured performance difference between data and simulation is applied to the simulation as a function of the \pt of the generated \PW and \PZ boson, following the procedure documented in~\cite{CMS:2011aa}.

The isolation requirements reduce the contributions from QCD multijet events. The remaining QCD multijet background contributions are estimated using control samples in observed data. A misidentification factor is defined and calculated as:
    \begin{equation}
        \mathrm{MF} = \frac{N^{\text{Iso}}_{\mathrm{QCD}}}{N^\text{Non-Iso}_{\mathrm{QCD}}}
    \end{equation}
where $N^{\text{Iso}}_\mathrm{QCD}$ and $N^{\text{Non-Iso}}_\mathrm{QCD}$ are the QCD multijet contributions in the lepton isolated and antiisolated regions, respectively. They are estimated by subtracting the contributions of prompt isolated leptons from observed data, estimated using EW simulation samples. The misidentification factors are measured separately for different flavors (\Pe and \PGm), binned in lepton \pt, $\eta$, and $\Delta\phi$ between the lepton and the neutrino directions. The $m_{\mathrm{T}}<20\GeV$ region is used to measure the misidentification factors; a background-dominated region with $20<m_{\mathrm{T}}<40\GeV$ confirms the validity of this procedure within the statistical uncertainties, which is around a few percent. The measured MFs are applied to the QCD multijet events in the antiisolated region with $m_\mathrm{T}>40\GeV$ to estimate the contribution in the \PW boson signal region.

The statistical uncertainties in the misidentification factors are propagated to the final fits and treated as uncertainties that modify the shape of the observable distributions, hereafter referred to as shape uncertainties. In addition, the systematic effects due to signal contamination in the lepton isolated and antiisolated regions are considered. Contributions from prompt isolated leptons estimated from the simulation, \ie, signal, EW background, and \ttbar processes, in the lepton (non)isolated regions are varied by 10 (20)\%, corresponding roughly to their contributions in these regions. The resulting $m_\mathrm{T}$ difference is included as a shape uncertainty in the maximum likelihood fit.

\begin{figure*}[!ht]
    \centering
    \includegraphics[width=0.49\textwidth]{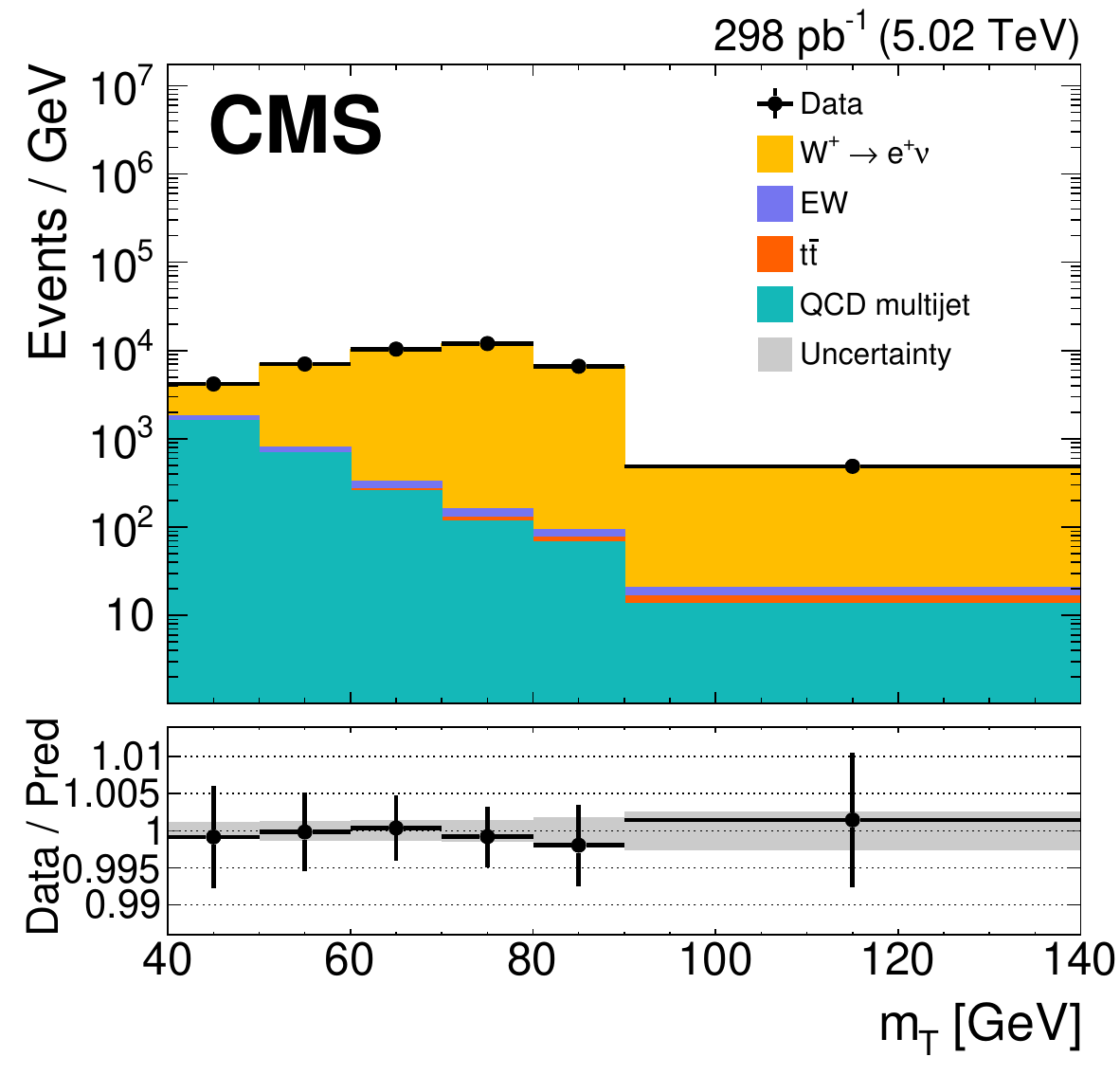}
    \includegraphics[width=0.49\textwidth]{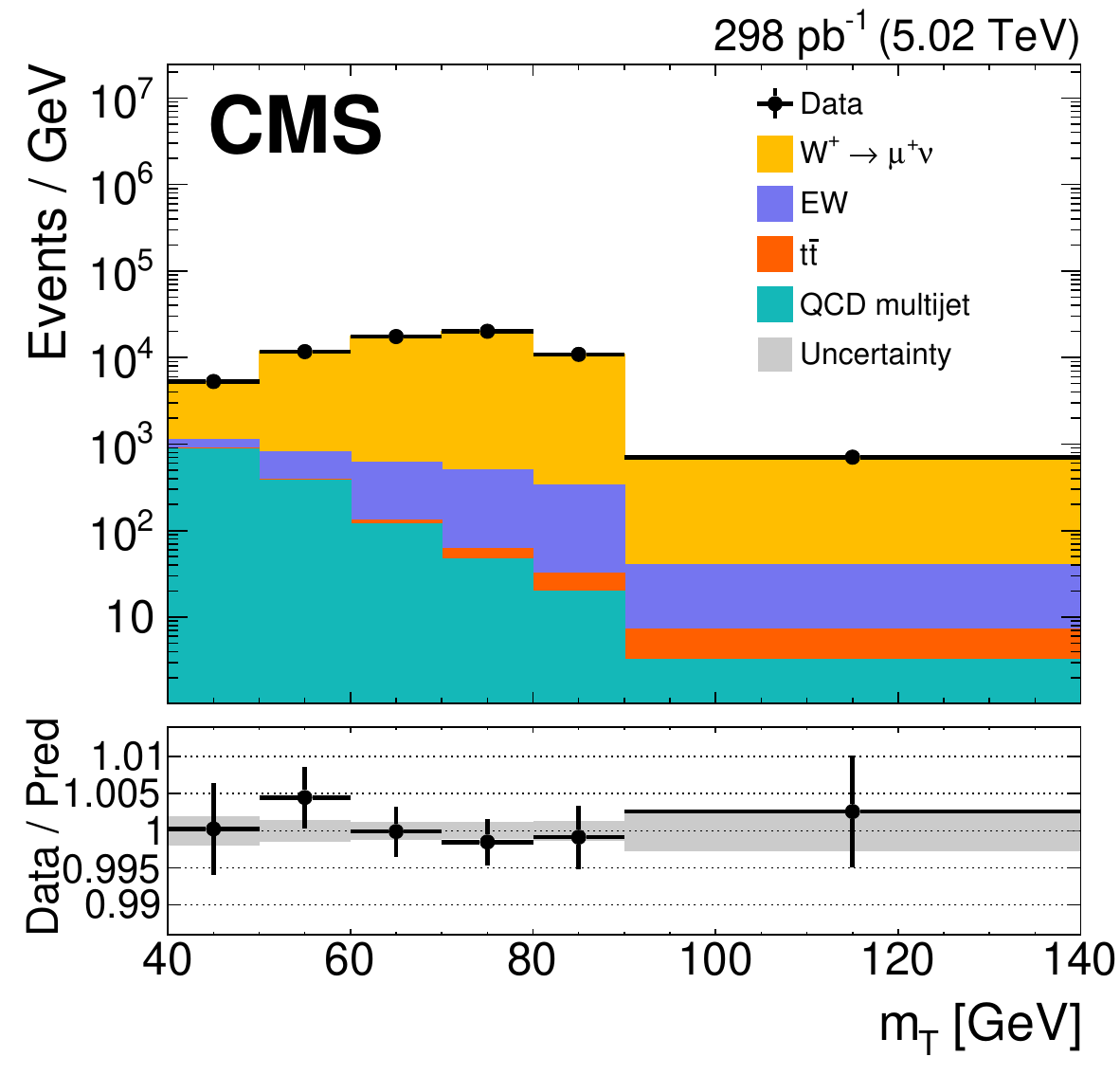}
    \includegraphics[width=0.49\textwidth]{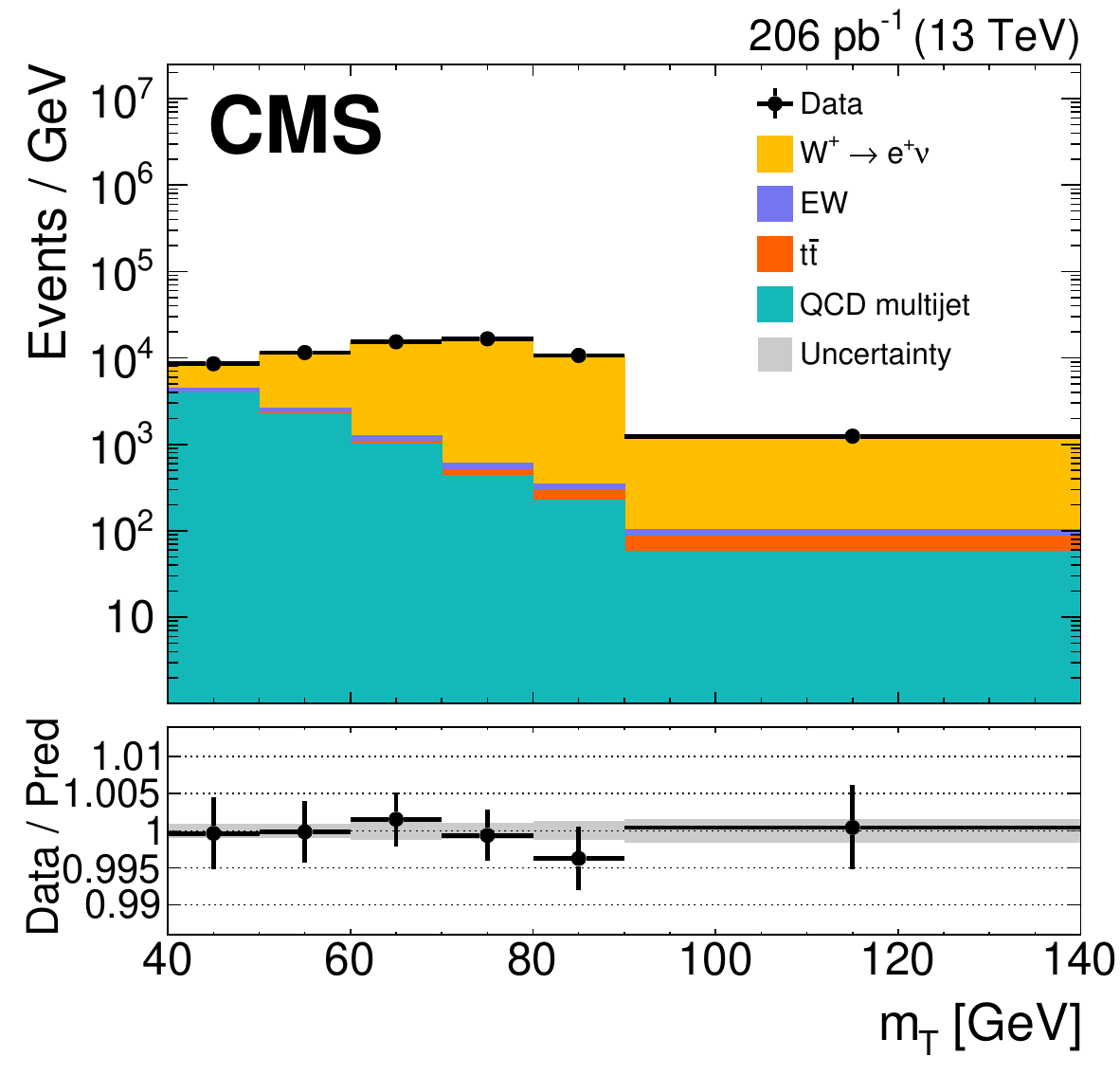}
    \includegraphics[width=0.49\textwidth]{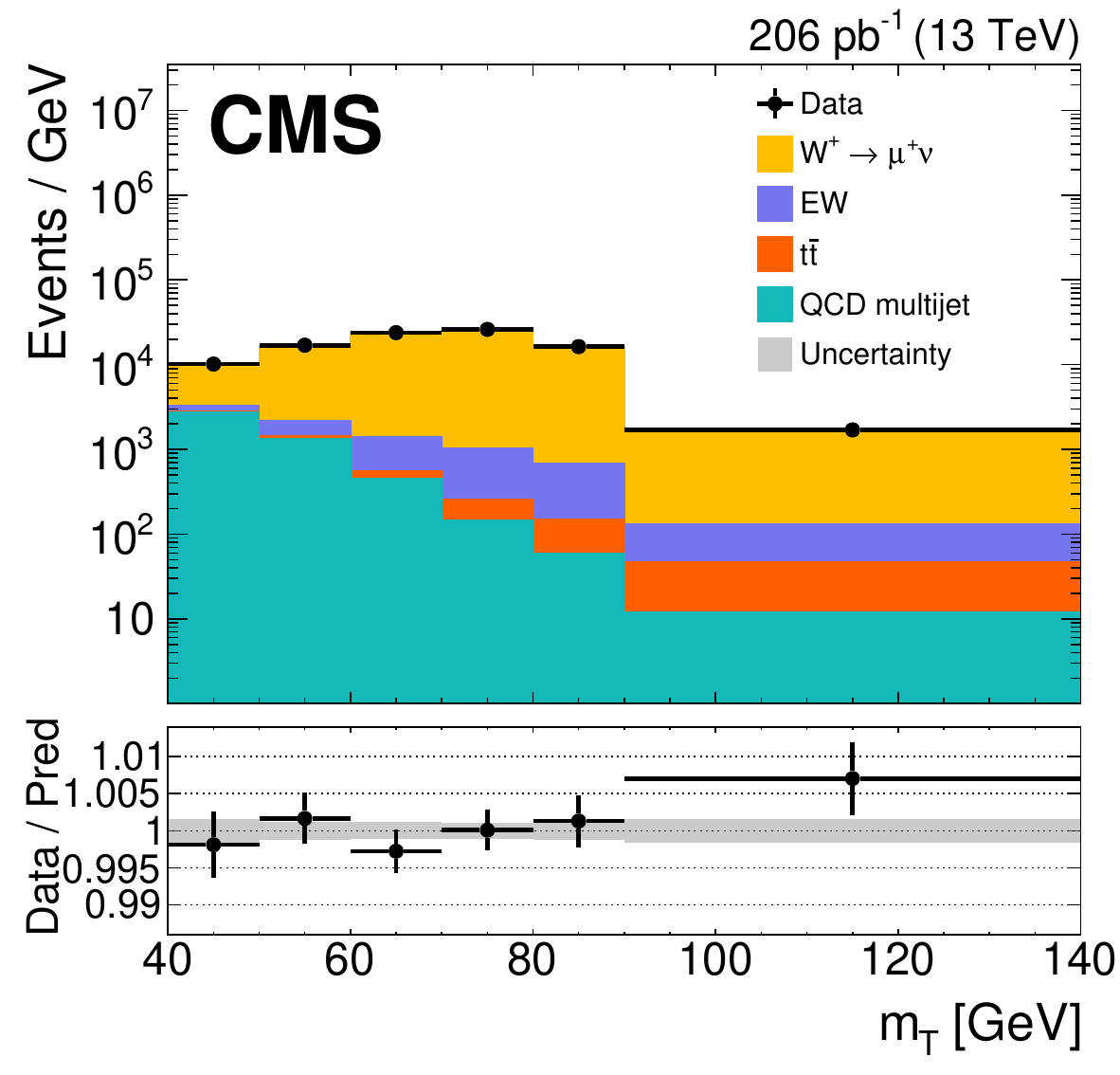}
    \caption{Distributions of $m_{\mathrm{T}}$ in the $\PWp$ signal selection for electron (left) and muon (right) final states for the $\Pp\Pp$ collisions at $\sqrt{s}=5.02\TeV$ (upper) and $13\TeV$ (lower) after the maximum likelihood fit. The vertical bars on the observed data represent corresponding statistical uncertainties. The EW backgrounds include the contributions from DY, $\PW\to\PGt\PGn$, and diboson processes. The lower panel in each plot shows the ratio of the number of observed events to that of the total signal and background predictions. The gray band represents the uncertainty in the total prediction after the fit.}
    \label{fig:signal_wp}
\end{figure*}

\begin{figure*}[!ht]
    \centering
    \includegraphics[width=0.49\textwidth]{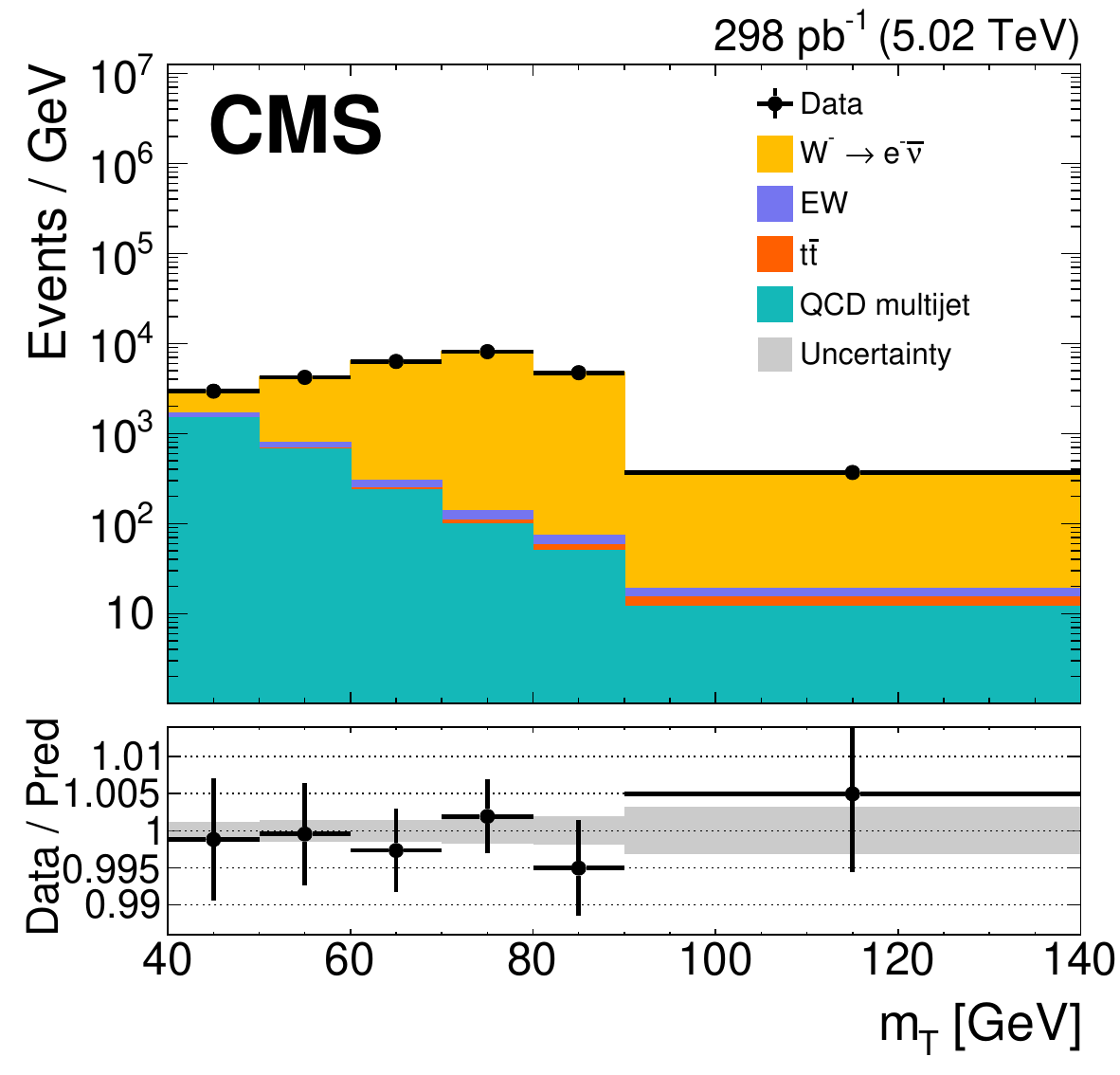}
    \includegraphics[width=0.49\textwidth]{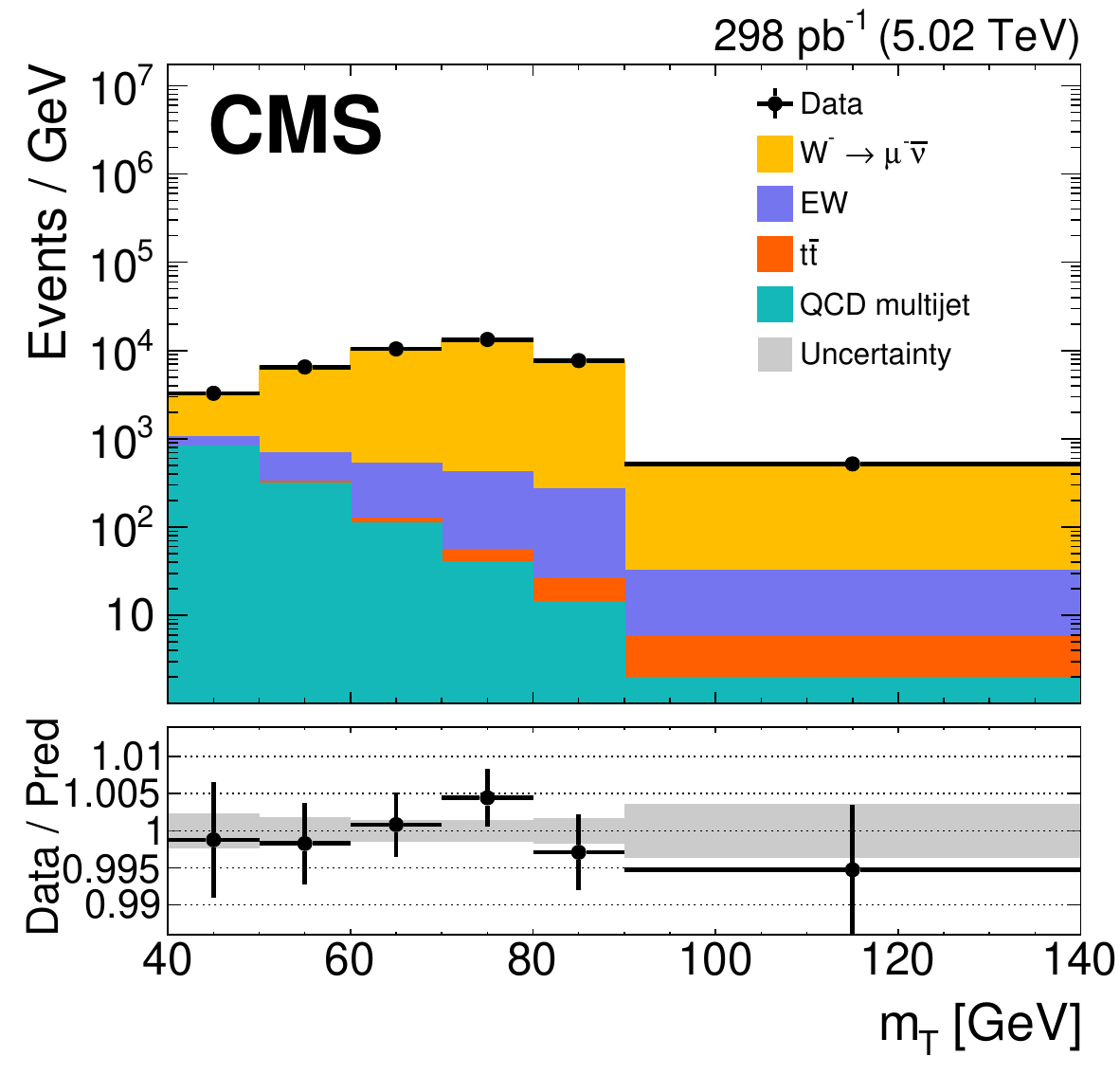}
    \includegraphics[width=0.49\textwidth]{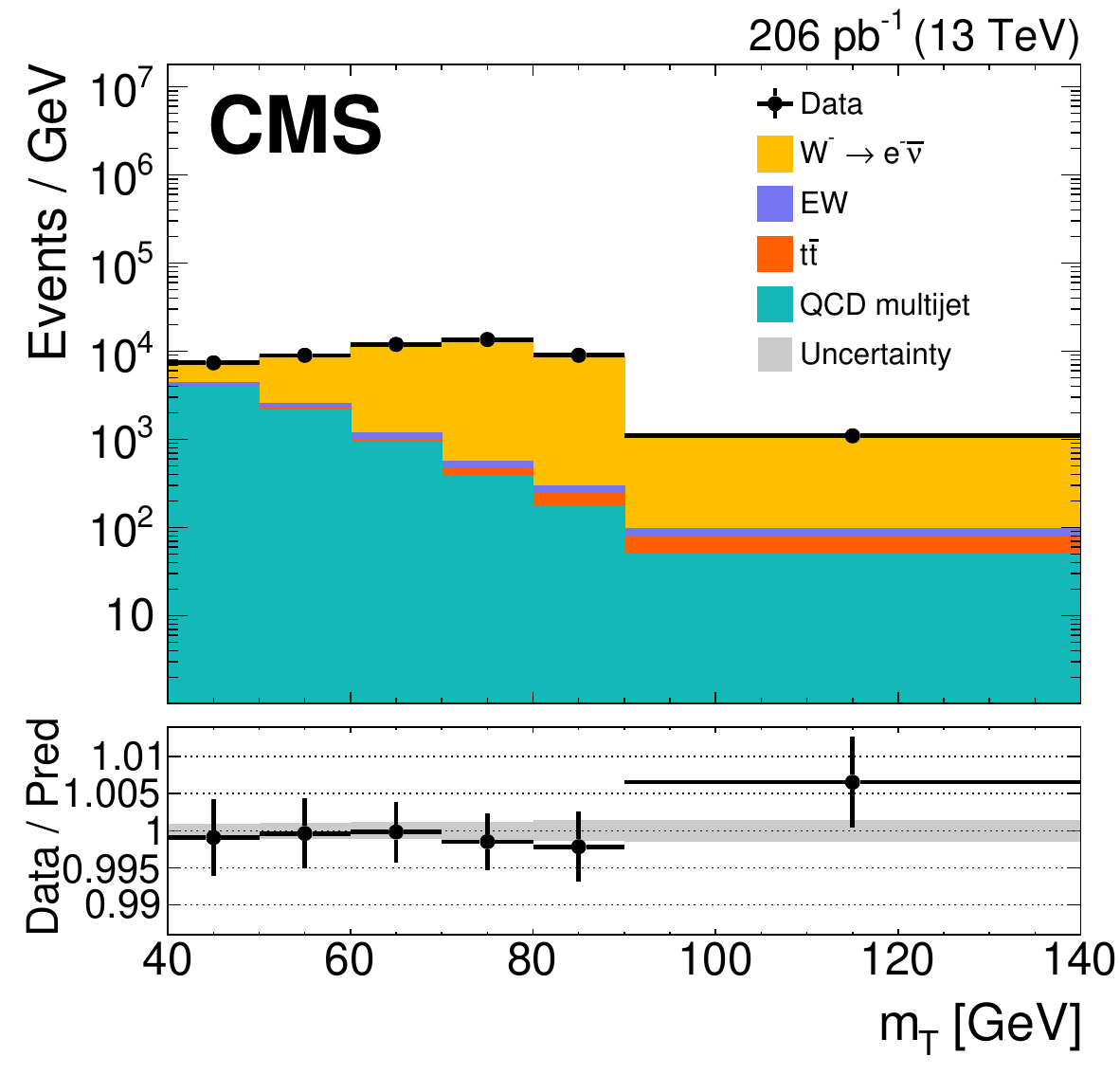}
    \includegraphics[width=0.49\textwidth]{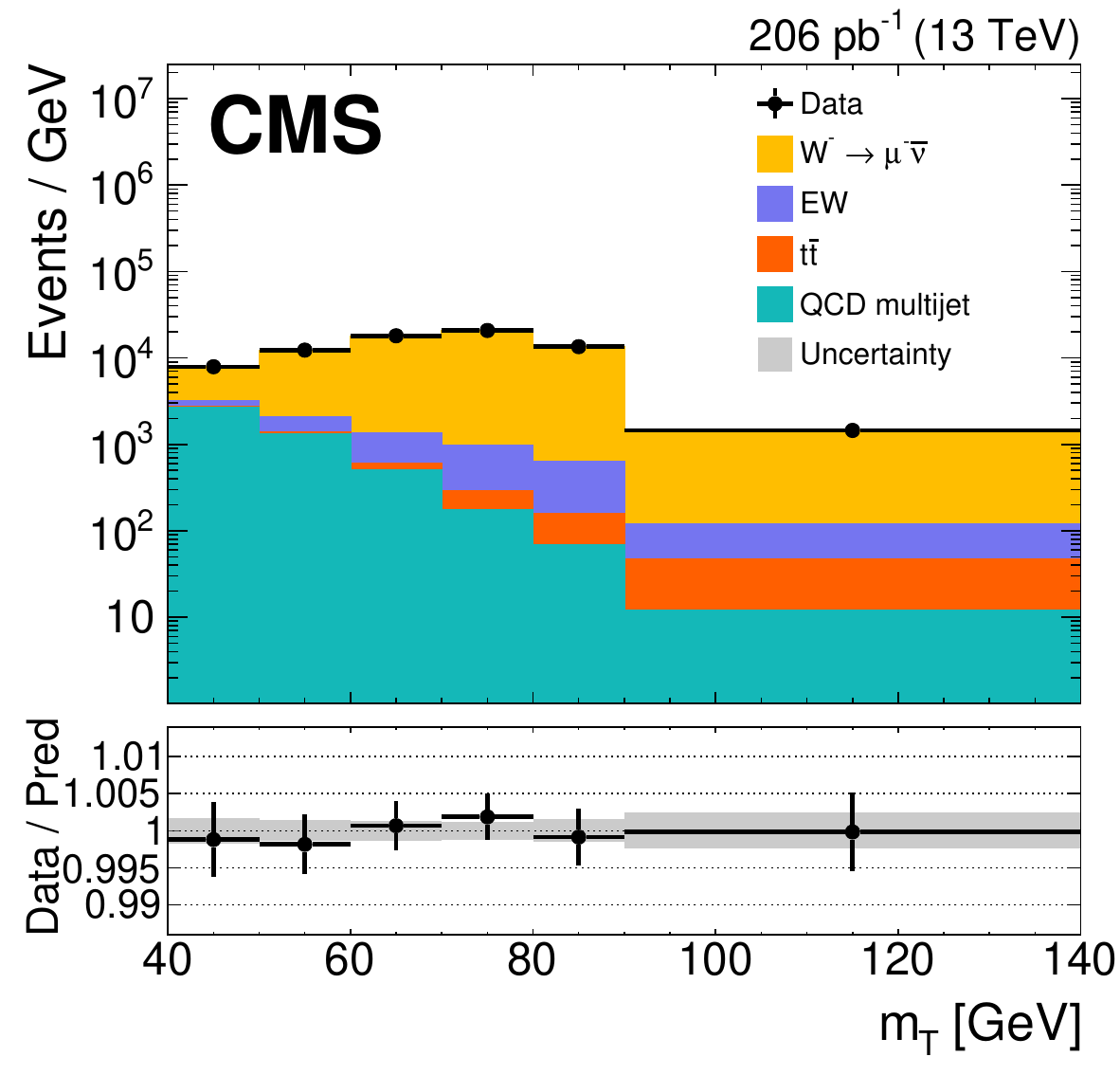}
    \caption{Distributions of $m_{\mathrm{T}}$ in the $\PWm$ signal selection for electron (left) and muon (right) final states for the $\Pp\Pp$ collisions at $\sqrt{s}=5.02\TeV$ (upper) and $13\TeV$ (lower) after the maximum likelihood fit. The EW backgrounds include the contributions from DY, $\PW\to\PGt\PGn$, and diboson processes. Notations are as in Fig.~\ref{fig:signal_wp}.}
    \label{fig:signal_wm}
\end{figure*}

\begin{figure*}[!ht]
    \centering
    \includegraphics[width=0.49\textwidth]{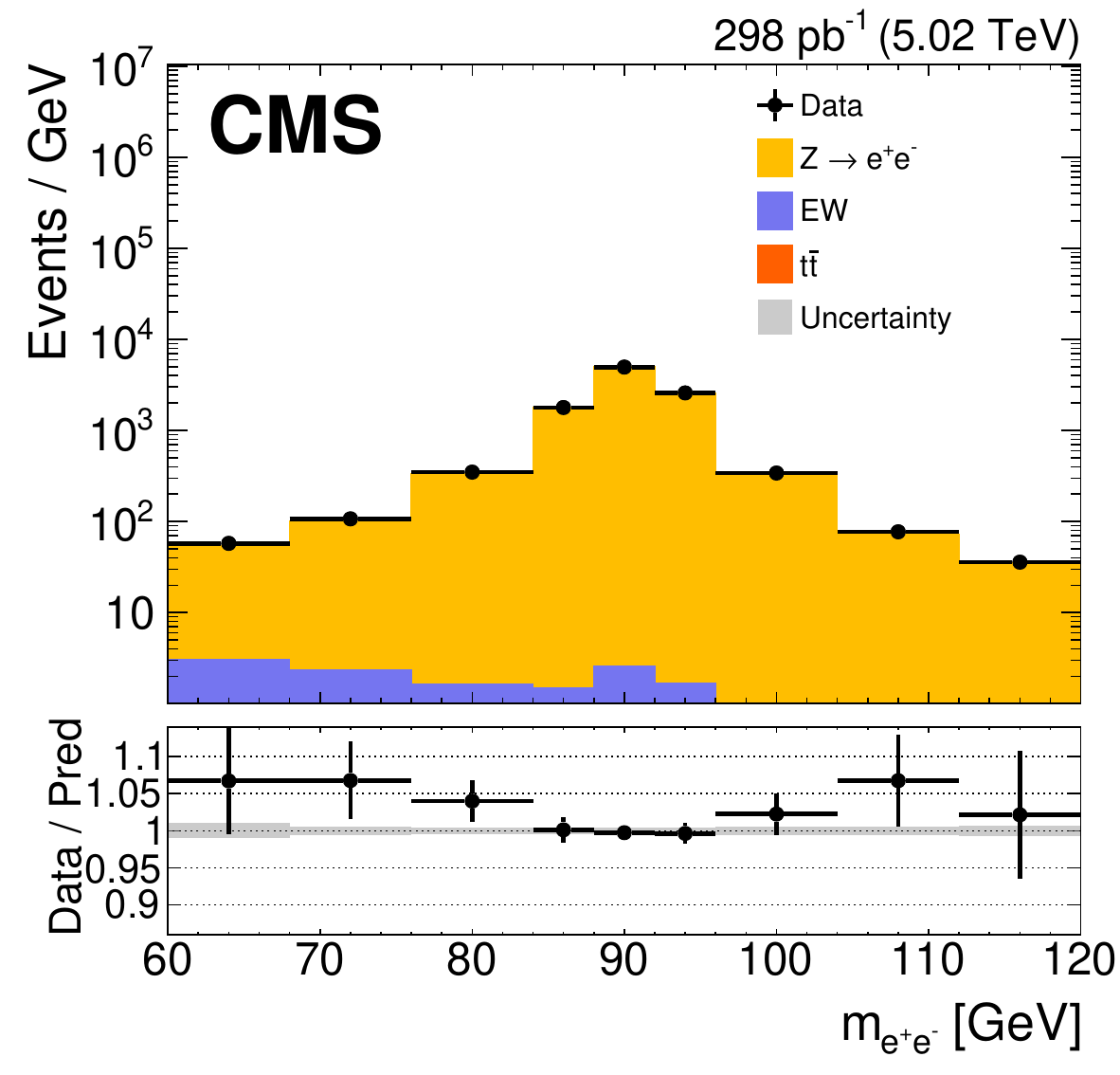}
    \includegraphics[width=0.49\textwidth]{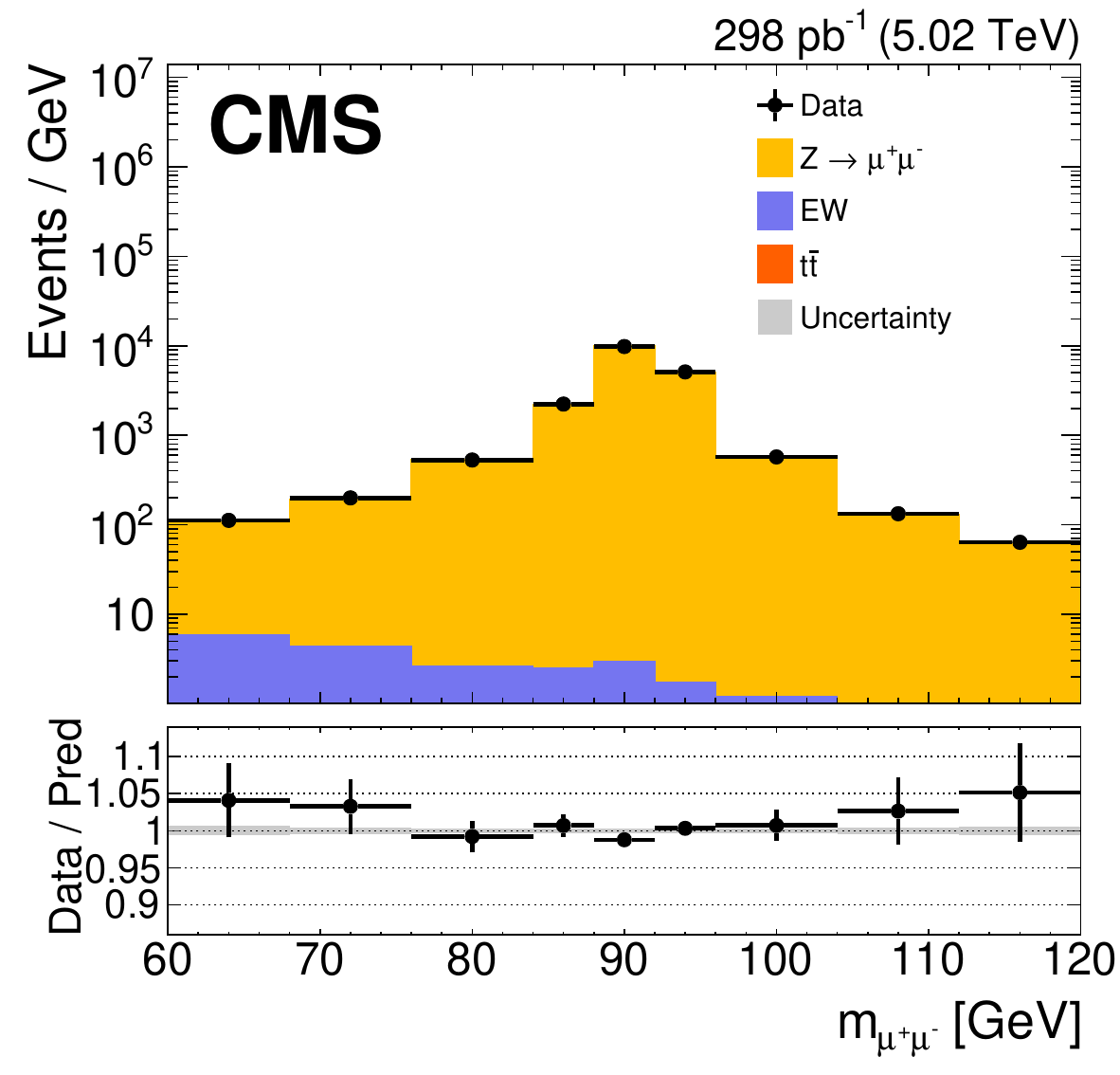}
    \includegraphics[width=0.49\textwidth]{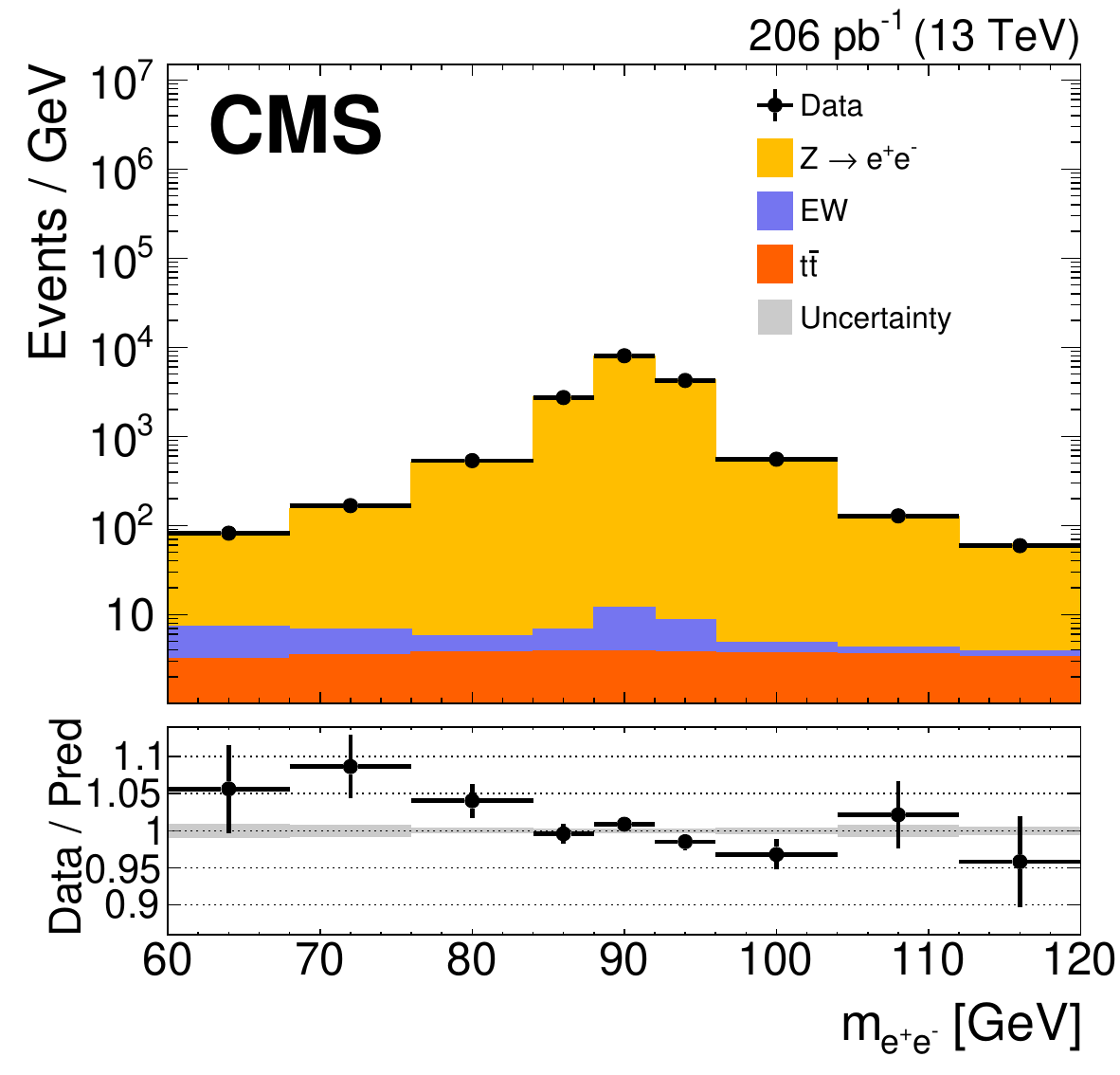}
    \includegraphics[width=0.49\textwidth]{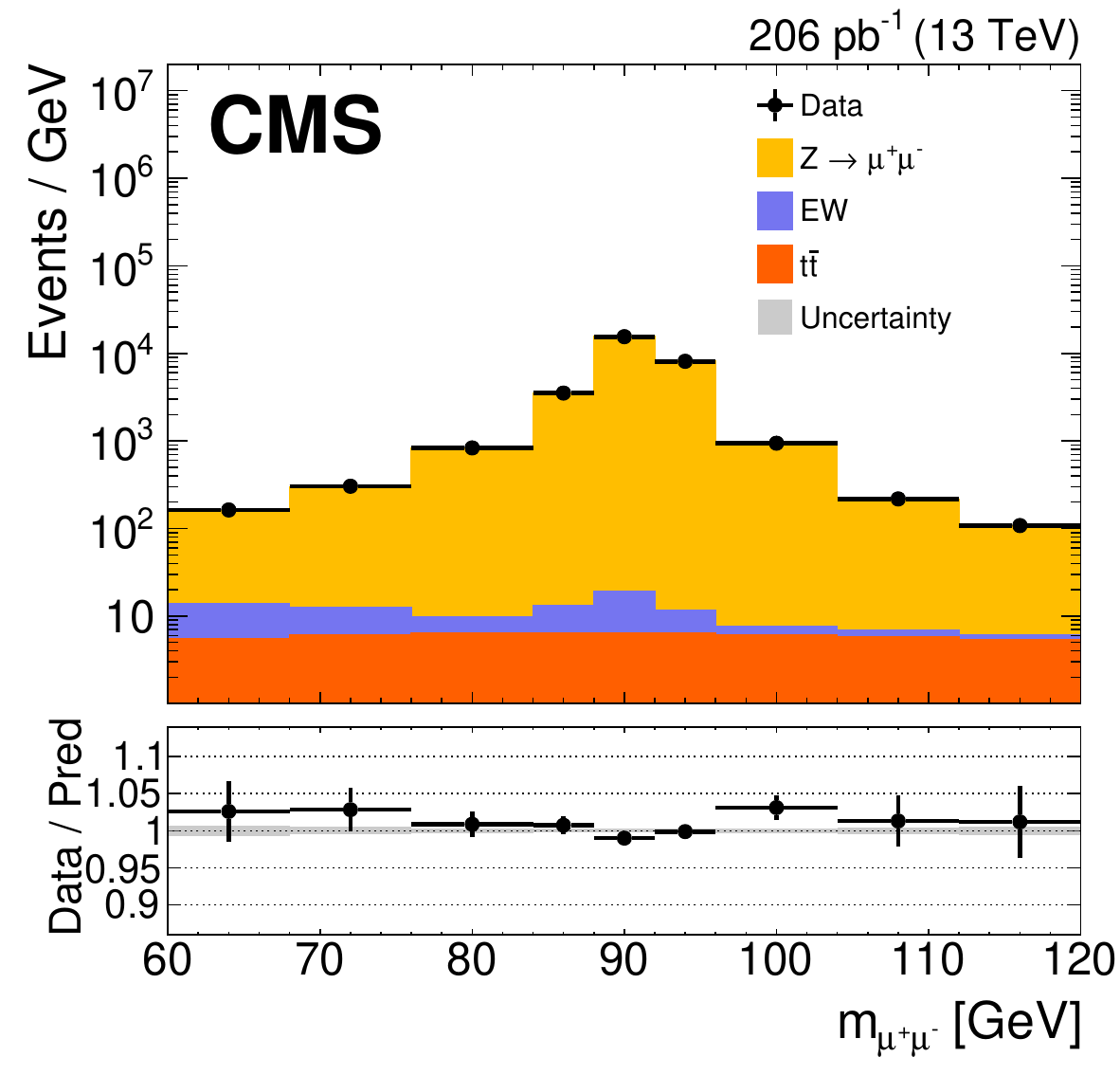}
    \caption{Distributions of $m_{\Pep\Pem}$ (left) and $m_{\PGmp\PGmm}$ (right) for the $\Pp\Pp$ collisions at $\sqrt{s}=5.02\TeV$ (upper) and $13\TeV$ (lower), after the maximum likelihood fit. The EW backgrounds include the contributions from diboson processes. Notations are as in Fig.~\ref{fig:signal_wp}.}
    \label{fig:signal_zll}
\end{figure*}

The distributions of $m_{\mathrm{T}}$ in the signal regions of the $\PWp$ ($\PWm$) boson in the electron and muon final states are shown in Fig.~\ref{fig:signal_wp} (\ref{fig:signal_wm}) at 5.02\TeV (upper) and 13\TeV (lower). The distributions of $m_{\Pep\Pem}$ and $m_{\PGmp\PGmm}$ in the signal region for the $\Pp\Pp$ collisions at $\sqrt{s}=5.02\TeV$ (upper plots) and $13\TeV$ (lower ones) are shown in Fig.~\ref{fig:signal_zll}. The predicted yields are shown with their best fit normalizations.  The data yields, together with the contributions of the different processes in the electron and muon final states at 5.02 and 13\TeV after the maximum likelihood fit, are shown in Tables~\ref{tab:nevts_5TeV} and \ref{tab:nevts_13TeV}, respectively. As a check of the methodology, fits were also performed switching to CT18~\cite{Hou:2019efy} PDFs at NNLO in QCD, and the difference on the results is negligible.

\begin{table}[!ht]
    \topcaption{Event yields, after the maximum likelihood fit, in the $\PWp$, $\PWm$, and \PZ boson signal regions for electron and muon final states at 5.02\TeV. The uncertainties include only the statistical component.}\label{tab:nevts_5TeV}
    \centering
    \newcolumntype{A}{r@{$\,\pm\,$}r}
    \cmsTable{
        \begin{tabular}{lAAAAAA}
            \hline
            {}           & \multicolumn{2}{c}{$\PWp\to\Pep\PGn$} & \multicolumn{2}{c}{$\PWm\to\Pem\PAGn$} & \multicolumn{2}{c}{$\PZ\to\Pep\Pem$} & \multicolumn{2}{c}{$\PWp\to\PGmp\PGn$} & \multicolumn{2}{c}{$\PWm\to\PGmm\PAGn$} & \multicolumn{2}{c}{$\PZ\to\PGmp\PGmm$}                                                                 \\
            \hline
            Observed     & \multicolumn{2}{c}{426023}                          & \multicolumn{2}{c}{281098}                           & \multicolumn{2}{c}{45031}                          & \multicolumn{2}{c}{688565}                           & \multicolumn{2}{c}{438044}                            & \multicolumn{2}{c}{81257}                                                                                            \\
            Signal       & 393080                                              & 630                                                  & 250360                                             & 500                                                  & 44720                                                 & 210                                                  & 652320 & 810   & 406600 & 640                   & 81310 & 290 \\
            EW           & 4610                                                & 68                                                   & 4224                                               & 65                                                   & 64.6                                                  & 8.0                                                  & 20640  & 140   & 17370  & 130                   & 107   & 10  \\
            $\ttbar$     & 572                                                 & 24                                                   & 577                                                & 24                                                   & 31.4                                                  & 5.6                                                  & 795    & 28    & 797    & 28                    & 50.5  & 7.1 \\
            QCD multijet & 27950                                               & 170                                                  & 26150                                              & 160                                                  & \multicolumn{2}{c}{\NA}                                 & 14600                                                & 120    & 13110 & 110    & \multicolumn{2}{c}{\NA}               \\
            \hline
        \end{tabular}
    }
\end{table}

\begin{table}[!ht]
    \topcaption{Event yields, after the maximum likelihood fit, in the $\PWp$, $\PWm$, and \PZ boson signal regions for electron and muon final states at 13\TeV. The uncertainties include only the statistical component.}\label{tab:nevts_13TeV}
    \newcolumntype{A}{r@{$\,\pm\,$}r}
    \cmsTable{
        \begin{tabular}{lAAAAAA}
            \hline
            {}           & \multicolumn{2}{c}{$\PWp\to\Pep\PGn$} & \multicolumn{2}{c}{$\PWm\to\Pem\PAGn$} & \multicolumn{2}{c}{$\PZ\to\Pep\Pem$} & \multicolumn{2}{c}{$\PWp\to\PGmp\PGn$} & \multicolumn{2}{c}{$\PWm\to\PGmm\PAGn$} & \multicolumn{2}{c}{$\PZ\to\PGmp\PGmm$}                                                                  \\
            \hline
            Observed     & \multicolumn{2}{c}{689131}                          & \multicolumn{2}{c}{561870}                           & \multicolumn{2}{c}{72040}                          & \multicolumn{2}{c}{1016318}                          & \multicolumn{2}{c}{796731}                            & \multicolumn{2}{c}{128889}                                                                                            \\
            Signal       & 591760                                              & 770                                                  & 467820                                             & 680                                                  & 71520                                                 & 270                                                  & 923620 & 960   & 708680 & 840                   & 128390 & 360 \\
            EW           & 12150                                               & 110                                                  & 11450                                              & 110                                                  & 159                                                   & 13                                                   & 38200  & 200   & 33710  & 180                   & 271    & 16  \\
            $\ttbar$     & 4768                                                & 69                                                   & 4780                                               & 69                                                   & 216                                                   & 15                                                   & 6326   & 80    & 6345   & 80                    & 360    & 19  \\
            QCD multijet & 80750                                               & 280                                                  & 77980                                              & 280                                                  & \multicolumn{2}{c}{\NA}                                 & 47910                                                & 220    & 47930 & 220    & \multicolumn{2}{c}{\NA}                \\
            \hline
        \end{tabular}
    }
\end{table}

\section{Systematic uncertainties}
\label{sec:systematics}
The sources of systematic uncertainty in the measurement include the
uncertainties in the integrated luminosity, lepton efficiencies
(reconstruction, identification, and trigger), prefiring, lepton momentum scale and resolution, background estimation, \ptmiss modeling, and the theoretical uncertainty in the kinematic acceptance. Different sources of systematic uncertainties can affect the rates and shapes of the $m_{\Pell^{+}\Pell^{-}}$ and $m_{\mathrm{T}}$ distributions for the signal and background processes in the simultaneous binned maximum likelihood fit, using the CMS statistical analysis tool \textsc{Combine}~\cite{CAT-23-001}.

The largest source of uncertainty in the inclusive cross section
measurements comes from the measurement of the integrated luminosity and
amounts to 1.9 and 2.3\% for the 5.02 and 13\TeV data samples~\cite{CMS-PAS-LUM-17-004,CMS-PAS-LUM-19-001}. The integrated luminosity uncertainty for the cross section ratios between 13 and 5.02\TeV is about 2.8\%. It gets canceled in the measurements of the cross section ratios at the same $\sqrt{s}$, and also in the double ratios between 13 and 5.02\TeV.

The systematic uncertainties in the measurement of the reconstruction, identification, and trigger efficiencies are estimated by studying the modeling of the background and signal parameterization in the $m_{\Pell^{+}\Pell^{-}}$ fit used in the tag-and-probe technique. The uncertainty in the modeling of the electromagnetic FSR in the tag-and-probe fits is obtained by weighting the simulation to reflect the differences between \PYTHIA and \textsc{photos} 3.56~\cite{Golonka:2005pn} modeling of the FSR with the exponentiation mode. An additional uncertainty is considered by varying the tag selection requirements in the efficiency measurement to estimate the bias coming from the tag selection in the tag-and-probe technique. The statistical uncertainties in the efficiency measurement are considered as uncorrelated among the lepton \pt, $\eta$, and charge bins, as they are measured separately and statistically independent.

Similar to the lepton efficiencies, the statistical and systematic prefiring uncertainties are also included. The systematic uncertainty is evaluated by varying the prefiring probability conservatively by $\pm20\%$, to cover the estimated uncertainty in this probability.

The uncertainties affecting the shape of the $m_{\mathrm{T}}$ distributions are considered with alternative shapes in the maximum likelihood fit. These include uncertainties in modeling the lepton momentum scale and resolution and also in calibrating the hadronic recoil of the \PW boson~\cite{Chatrchyan:2014mua}. The statistical uncertainties in the recoil calibrations are propagated to the final fits. The modeling of the recoil distributions is also varied, by changing the fitting functions and varying the binning choice. The difference between variations is treated as the systematic uncertainty. These uncertainty contributions are small.

A normalization uncertainty of 5\% is assigned to the top quark and diboson background processes to cover for any unaccounted mismodeling of the simulation. Because these processes are suppressed by the event selection, the effect of these uncertainties in the result is very small. The systematic uncertainty in the QCD multijet background estimation using control samples in observed data was described in Section~\ref{sec:backgrounds}. The statistical uncertainties due to the finite size of simulated samples are also included~\cite{BARLOW1993219}. The uncertainty due to the limited size of antiisolated control regions is parametrized with the uncertainty of MF as described in Section~\ref{sec:backgrounds}.

The theoretical uncertainties in the kinematic acceptance are estimated by varying the renormalization ($\mu_{\mathrm{R}}$) and factorization ($\mu_{\mathrm{F}}$) scales independently up and down by a factor of two from their nominal values (excluding the variations where $\mu_\mathrm{R}$ and $\mu_\mathrm{F}$ scales are set to lowest and highest values at the same time) and taking the largest acceptance variation as the uncertainty. The uncertainties from the NNPDF~3.1 PDF set and the strong coupling \alpS are evaluated according to the procedure described in Ref.~\cite{Butterworth:2015oua}.
The fixed-order calculations are unreliable at low vector boson \pt due to soft and collinear gluon radiation, resulting in large logarithmic corrections~\cite{Collins:1984kg}. Fixed-order perturbative calculations combined with parton shower models~\cite{Sjostrand:2014zea,Gleisberg:2008ta,Bahr:2008pv} provide fully exclusive predictions~\cite{Nason:2004rx,Frixione:2002ik,Alioli:2010xd,Alwall:2014hca} with only leading logarithmic formal accuracy. Resummation of the logarithmically divergent terms at next-to-next-to-leading logarithmic (NNLL) accuracy has been matched with the fixed-order predictions to achieve accurate predictions for the entire boson \pt range~\cite{Balazs:1995nz,Catani:2015vma}.
The differences in the kinematic acceptance between the nominal \MGvATNLO and resummed \textsc{DYturbo}~\cite{Camarda:2019zyx,Camarda:2021ict,Camarda:2021jsw,Camarda:2023dqn} predictions are taken as a systematic uncertainty. Finally, the uncertainty in the modeling of the FSR is obtained by comparing the acceptance differences between \PYTHIA  and \textsc{photos} predictions.

The systematic uncertainties in the fiducial inclusive cross sections and ratios at 5.02 and 13\TeV are summarized in Tables~\ref{tab:syst:5} and~\ref{tab:syst:13}, respectively, for both the electron and muon final states. Table~\ref{tab:syst:13vs5} includes the systematic uncertainties in the fiducial inclusive cross section ratios between 13 and 5.02\TeV.

Similar systematic uncertainties for the total inclusive cross sections and ratios are summarized in Tables~\ref{tab:syst:5inc},~\ref{tab:syst:13inc}, and~\ref{tab:syst:13vs5inc}, where the theoretical uncertainties in the acceptance are listed.

\begin{table}[htbp]
    \centering
    \topcaption{Systematic uncertainties, in percent, for the fiducial inclusive cross sections at 5.02\TeV. The 1.9\% integrated luminosity uncertainty, which affects the \PW and \PZ boson production cross sections, is not included in the table.}
    \cmsTable{
        \begin{tabular}{lcccccc}
            \hline
            {}                                               & $\PWp\to \Pell^{+}\PGn$ & $\PWm\to \Pell^{-}\PAGn$ & $\PW\to \Pell\PGn$ & $\PZ\to \Pell^{+}\Pell^{-}$ & $\PWp/\PWm$ & $\PW/\PZ$ \\
            \hline
            Total                                            & 0.32                               & 0.34                                & 0.27                       & 0.37                                & 0.40              & 0.25      \\
            Efficiency (stat)                                & 0.24                               & 0.22                                & 0.17                       & 0.27                                & 0.30              & 0.11      \\
            Trigger prefire correction                       & 0.15                               & 0.14                                & 0.14                       & 0.22                                & 0.01              & 0.08      \\
            QCD multijet (syst)                              & 0.11                               & 0.14                                & 0.09                       & 0.11                                & 0.18              & 0.15      \\
            MC sim. stat                                     & 0.10                               & 0.12                                & 0.08                       & 0.11                                & 0.15              & 0.12      \\
            EW + $\ttbar$ cross section                      & 0.08                               & 0.11                                & 0.10                       & 0.02                                & 0.03              & 0.08      \\
            $\mu_{\mathrm{R}}$ and $\mu_{\mathrm{F}}$ scales & 0.06                               & 0.07                                & 0.06                       & 0.02                                & 0.02              & 0.05      \\
            Efficiency (syst)                                & 0.04                               & 0.05                                & 0.04                       & 0.09                                & 0.01              & 0.06      \\
            PDF + $\alpha_\mathrm{S}$                        & 0.04                               & 0.06                                & 0.03                       & 0.02                                & 0.07              & 0.05      \\
            Hadronic recoil calibration                      & 0.03                               & 0.04                                & 0.04                       & 0.01                                & 0.02              & 0.05      \\
            QCD multijet (stat)                              & 0.03                               & 0.05                                & 0.03                       & 0.03                                & 0.05              & 0.04      \\
            \hline
        \end{tabular}
    }
    \label{tab:syst:5}
\end{table}

\begin{table}[htbp]
    \topcaption{Systematic uncertainties, in percent, for the fiducial inclusive cross sections at 13\TeV. The 2.3\% integrated luminosity uncertainty, which affects the \PW and \PZ boson production cross sections, is not included in the table.}
    \centering
    \cmsTable{
        \begin{tabular}{lcccccc}
            \hline
            {}                                               & $\PWp\to \Pell^{+}\PGn$ & $\PWm\to \Pell^{-}\PAGn$ & $\PW\to \Pell\PGn$ & $\PZ\to \Pell^{+}\Pell^{-}$ & $\PWp/\PWm$ & $\PW/\PZ$ \\
            \hline
            Total                                            & 0.35                               & 0.35                                & 0.29                       & 0.40                                & 0.40              & 0.27      \\
            Efficiency (stat)                                & 0.23                               & 0.21                                & 0.17                       & 0.26                                & 0.29              & 0.11      \\
            Trigger prefire correction                       & 0.17                               & 0.16                                & 0.17                       & 0.26                                & 0.01              & 0.09      \\
            QCD multijet (syst)                              & 0.16                               & 0.16                                & 0.11                       & 0.14                                & 0.25              & 0.17      \\
            EW + $\ttbar$ cross section                      & 0.10                               & 0.11                                & 0.10                       & 0.01                                & 0.01              & 0.10      \\
            MC sim. stat                                     & 0.10                               & 0.11                                & 0.08                       & 0.12                                & 0.13              & 0.12      \\
            Efficiency (syst)                                & 0.08                               & 0.09                                & 0.08                       & 0.15                                & 0.01              & 0.08      \\
            PDF + $\alpha_\mathrm{S}$                        & 0.08                               & 0.07                                & 0.05                       & 0.03                                & 0.11              & 0.06      \\
            $\mu_{\mathrm{R}}$ and $\mu_{\mathrm{F}}$ scales & 0.07                               & 0.07                                & 0.07                       & 0.07                                & 0.03              & 0.05      \\
            QCD multijet (stat)                              & 0.02                               & 0.03                                & 0.02                       & 0.02                                & 0.04              & 0.02      \\
            Hadronic recoil calibration                      & 0.02                               & 0.02                                & 0.02                       & 0.02                                & 0.01              & 0.02      \\
            \hline
        \end{tabular}
    }
    \label{tab:syst:13}
\end{table}

\begin{table}[htbp]
    \topcaption{Systematic uncertainties, in percent, for the fiducial inclusive cross section ratios between 13 and 5.02\TeV. The 2.8\% integrated luminosity uncertainty, which affects the \PW and \PZ boson production cross section ratios, is not included in the table.}
    \centering
    \cmsTable{
        \begin{tabular}{lcccccc}
            \hline
            {}                                               & $\PWp\to \Pell^{+}\PGn$ & $\PWm\to \Pell^{-}\PAGn$ & $\PW\to \Pell\PGn$ & $\PZ\to \Pell^{+}\Pell^{-}$ & $\PWp/\PWm$ & $\PW/\PZ$ \\
            \hline
            Total                                            & 0.48                               & 0.49                                & 0.39                       & 0.55                                & 0.56              & 0.36      \\
            Efficiency (stat)                                & 0.33                               & 0.30                                & 0.24                       & 0.37                                & 0.41              & 0.16      \\
            Trigger prefire correction                       & 0.23                               & 0.21                                & 0.22                       & 0.34                                & 0.02              & 0.12      \\
            QCD multijet (syst)                              & 0.19                               & 0.21                                & 0.13                       & 0.18                                & 0.30              & 0.23      \\
            MC sim. stat                                     & 0.14                               & 0.17                                & 0.12                       & 0.17                                & 0.20              & 0.17      \\
            EW + $\ttbar$ cross section                      & 0.13                               & 0.15                                & 0.14                       & 0.02                                & 0.03              & 0.12      \\
            Efficiency (syst)                                & 0.09                               & 0.10                                & 0.09                       & 0.18                                & 0.01              & 0.10      \\
            $\mu_{\mathrm{R}}$ and $\mu_{\mathrm{F}}$ scales & 0.09                               & 0.10                                & 0.09                       & 0.08                                & 0.04              & 0.07      \\
            PDF + $\alpha_\mathrm{S}$                        & 0.07                               & 0.06                                & 0.04                       & 0.03                                & 0.09              & 0.04      \\
            Hadronic recoil calibration                      & 0.04                               & 0.05                                & 0.04                       & 0.02                                & 0.02              & 0.05      \\
            QCD multijet (stat)                              & 0.04                               & 0.05                                & 0.03                       & 0.04                                & 0.06              & 0.04      \\
            \hline
        \end{tabular}
    }
    \label{tab:syst:13vs5}
\end{table}

\begin{table}[htbp]
    \centering
    \topcaption{Systematic uncertainties, in percent, for the total inclusive cross sections at 5.02\TeV. The 1.9\% integrated luminosity uncertainty, which affects the \PW and \PZ boson production cross sections, is not included in the table.}
    \cmsTable{
        \begin{tabular}{lcccccc}
            \hline
            {}                                               & $\PWp\to \Pell^{+}\PGn$ & $\PWm\to \Pell^{-}\PAGn$ & $\PW\to \Pell\PGn$ & $\PZ\to \Pell^{+}\Pell^{-}$ & $\PWp/\PWm$ & $\PW/\PZ$ \\
            \hline
            Total                                            & 0.81                               & 0.89                                & 0.79                       & 0.82                                & 0.66              & 0.49      \\
            $\mu_{\mathrm{R}}$ and $\mu_{\mathrm{F}}$ scales & 0.67                               & 0.73                                & 0.70                       & 0.61                                & 0.08              & 0.13      \\
            Resum. + FSR                                     & 0.30                               & 0.37                                & 0.24                       & 0.35                                & 0.47              & 0.39      \\
            Efficiency (stat)                                & 0.23                               & 0.22                                & 0.16                       & 0.26                                & 0.31              & 0.11      \\
            Trigger prefire correction                       & 0.14                               & 0.13                                & 0.14                       & 0.25                                & 0.01              & 0.11      \\
            QCD multijet (syst)                              & 0.13                               & 0.16                                & 0.10                       & 0.13                                & 0.22              & 0.19      \\
            PDF + $\alpha_\mathrm{S}$                        & 0.13                               & 0.17                                & 0.10                       & 0.18                                & 0.22              & 0.12      \\
            MC sim. stat                                     & 0.12                               & 0.16                                & 0.12                       & 0.13                                & 0.16              & 0.14      \\
            EW + $\ttbar$ cross section                      & 0.07                               & 0.09                                & 0.08                       & 0.01                                & 0.02              & 0.07      \\
            QCD multijet (stat)                              & 0.06                               & 0.06                                & 0.06                       & 0.06                                & 0.06              & 0.05      \\
            Hadronic recoil calibration                      & 0.04                               & 0.05                                & 0.04                       & 0.05                                & 0.01              & 0.02      \\
            Efficiency (syst)                                & 0.04                               & 0.04                                & 0.04                       & 0.09                                & 0.02              & 0.06      \\
            \hline
        \end{tabular}
    }
    \label{tab:syst:5inc}
\end{table}

\begin{table}[htbp]
    \topcaption{Systematic uncertainties, in percent, for the total inclusive cross sections at 13\TeV. The 2.3\% integrated luminosity uncertainty, which affects the \PW and \PZ boson production cross sections, is not included in the table.}
    \centering
    \cmsTable{
        \begin{tabular}{lcccccc}
            \hline
            {}                                               & $\PWp\to \Pell^{+}\PGn$ & $\PWm\to \Pell^{-}\PAGn$ & $\PW\to \Pell\PGn$ & $\PZ\to \Pell^{+}\Pell^{-}$ & $\PWp/\PWm$ & $\PW/\PZ$ \\
            \hline
            Total                                            & 0.89                               & 0.92                                & 0.84                       & 0.90                                & 0.69              & 0.79      \\
            $\mu_{\mathrm{R}}$ and $\mu_{\mathrm{F}}$ scales & 0.62                               & 0.64                                & 0.63                       & 0.66                                & 0.08              & 0.67      \\
            PDF + $\alpha_\mathrm{S}$                        & 0.40                               & 0.45                                & 0.41                       & 0.43                                & 0.23              & 0.22      \\
            Resum. + FSR                                     & 0.40                               & 0.37                                & 0.29                       & 0.12                                & 0.51              & 0.26      \\
            Trigger prefire correction                       & 0.27                               & 0.25                                & 0.26                       & 0.34                                & 0.01              & 0.09      \\
            Efficiency (stat)                                & 0.25                               & 0.23                                & 0.19                       & 0.28                                & 0.29              & 0.11      \\
            QCD multijet (syst)                              & 0.18                               & 0.18                                & 0.10                       & 0.14                                & 0.30              & 0.20      \\
            MC sim. stat                                     & 0.15                               & 0.16                                & 0.14                       & 0.16                                & 0.16              & 0.15      \\
            Efficiency (syst)                                & 0.11                               & 0.11                                & 0.11                       & 0.16                                & 0.01              & 0.09      \\
            EW + $\ttbar$ cross section                      & 0.10                               & 0.12                                & 0.11                       & 0.04                                & 0.02              & 0.09      \\
            QCD multijet (stat)                              & 0.08                               & 0.09                                & 0.08                       & 0.07                                & 0.09              & 0.08      \\
            Hadronic recoil calibration                      & 0.02                               & 0.03                                & 0.02                       & 0.05                                & 0.02              & 0.03      \\
            \hline
        \end{tabular}
    }
    \label{tab:syst:13inc}
\end{table}

\begin{table}[htbp]
    \topcaption{Systematic uncertainties, in percent, for the total inclusive cross section ratios between 13 and 5.02\TeV. The 2.8\% integrated luminosity uncertainty, which affects the \PW and \PZ boson production cross section ratios, is not included in the table.}
    \centering
    \cmsTable{
        \begin{tabular}{lcccccc}
            \hline
            {}                                               & $\PWp\to \Pell^{+}\PGn$ & $\PWm\to \Pell^{-}\PAGn$ & $\PW\to \Pell\PGn$ & $\PZ\to \Pell^{+}\Pell^{-}$ & $\PWp/\PWm$ & $\PW/\PZ$ \\
            \hline
            Total                                            & 1.20                               & 1.31                                & 1.16                       & 1.22                                & 0.98              & 0.92      \\
            $\mu_{\mathrm{R}}$ and $\mu_{\mathrm{F}}$ scales & 0.91                               & 0.97                                & 0.94                       & 0.90                                & 0.11              & 0.68      \\
            Resum. + FSR                                     & 0.50                               & 0.52                                & 0.37                       & 0.37                                & 0.70              & 0.47      \\
            PDF + $\alpha_\mathrm{S}$                        & 0.41                               & 0.54                                & 0.44                       & 0.48                                & 0.40              & 0.20      \\
            Efficiency (stat)                                & 0.34                               & 0.32                                & 0.25                       & 0.39                                & 0.43              & 0.16      \\
            Trigger prefire correction                       & 0.30                               & 0.29                                & 0.30                       & 0.43                                & 0.02              & 0.14      \\
            QCD multijet (syst)                              & 0.22                               & 0.23                                & 0.13                       & 0.19                                & 0.37              & 0.27      \\
            MC sim. stat                                     & 0.19                               & 0.23                                & 0.18                       & 0.21                                & 0.23              & 0.21      \\
            EW + $\ttbar$ cross section                      & 0.12                               & 0.14                                & 0.13                       & 0.04                                & 0.02              & 0.11      \\
            Efficiency (syst)                                & 0.12                               & 0.12                                & 0.12                       & 0.18                                & 0.03              & 0.11      \\
            QCD multijet (stat)                              & 0.10                               & 0.11                                & 0.09                       & 0.09                                & 0.11              & 0.10      \\
            Hadronic recoil calibration                      & 0.04                               & 0.05                                & 0.05                       & 0.06                                & 0.02              & 0.04      \\
            \hline
        \end{tabular}
    }
    \label{tab:syst:13vs5inc}
\end{table}

For comparisons, the theoretical predictions of the fiducial and total inclusive cross sections and their ratios are computed at NNLO+NNLL in QCD with \textsc{DYturbo} v1.3.2 using the NNPDF 3.1~\cite{NNPDF:2017mvq}, NNPDF 4.0~\cite{NNPDF:2021njg}, CT18~\cite{Hou:2019efy}, and MSHT20~\cite{Bailey:2020ooq} PDF sets at NNLO in QCD. The uncertainties in these predictions, at 68\% confidence level, include contributions from statistical uncertainty, and uncertainties in the \alpS, PDFs, and $\mu_{\mathrm{R}}$ and $\mu_{\mathrm{F}}$. The $\mu_{\mathrm{R}}$ and $\mu_{\mathrm{F}}$ are varied together up and down by a factor of two from their nominal values and the envelope is taken as the uncertainty. The \PZ boson production cross sections require $m_{\Pell^{+}\Pell^{-}}$ within the range of 60--120\GeV, and include the effect of virtual photons.

\section{Results}
\label{sec:results}
\begin{figure}[htbp]
    \centering
    \includegraphics[width=0.99\textwidth]{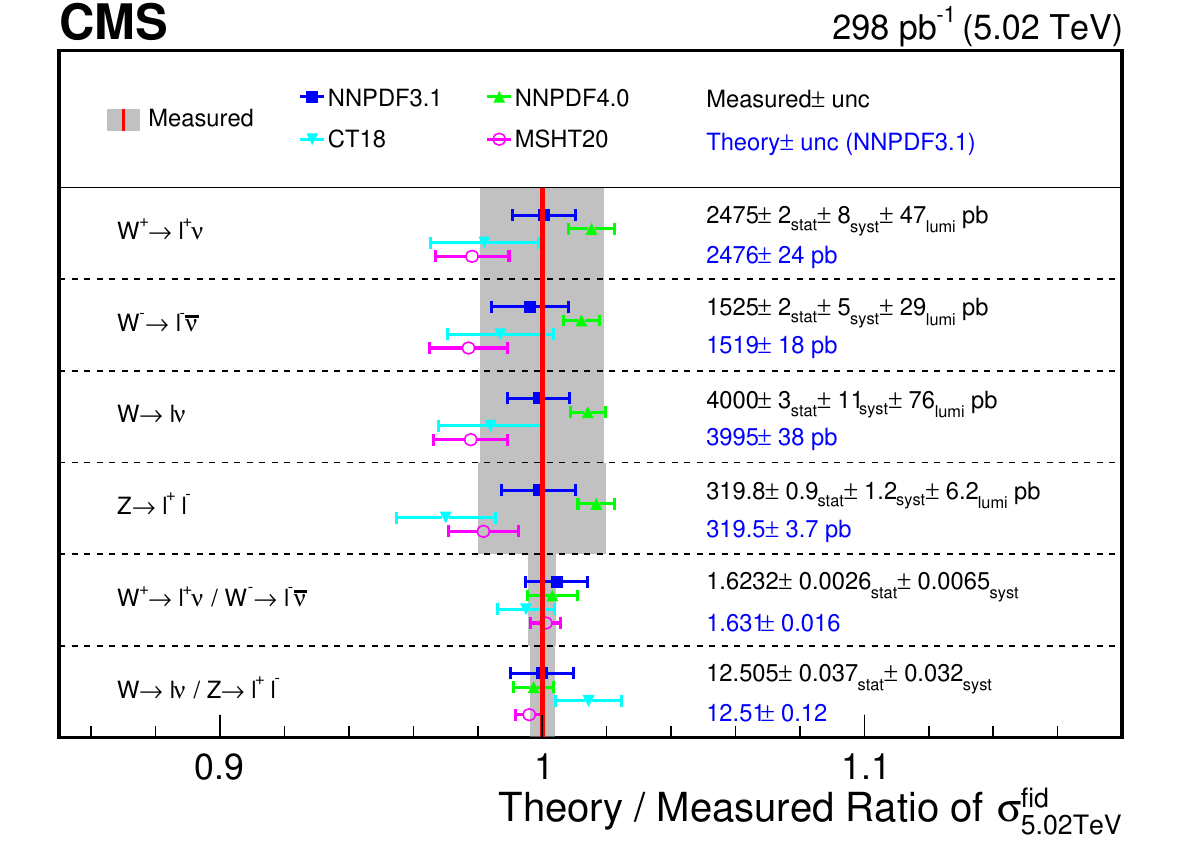}
    \caption{Comparisons of the fiducial inclusive cross sections and cross section ratios between measurements and the theoretical calculations from \textsc{DYturbo} with different PDF sets at 5.02\TeV. The gray band represents the total uncertainty of each measurement. The uncertainties in the theoretical predictions include the statistical uncertainty, and the PDF, \alpS, and renormalization and factorization scale uncertainties. The measured values and theoretical predictions (\textsc{DYturbo} with NNPDF 3.1 as the example) are also shown in the right part of the plot.}
    \label{fig:xsec_fid_5TeV}
\end{figure}

\begin{figure}[htbp]
    \centering
    \includegraphics[width=0.99\textwidth]{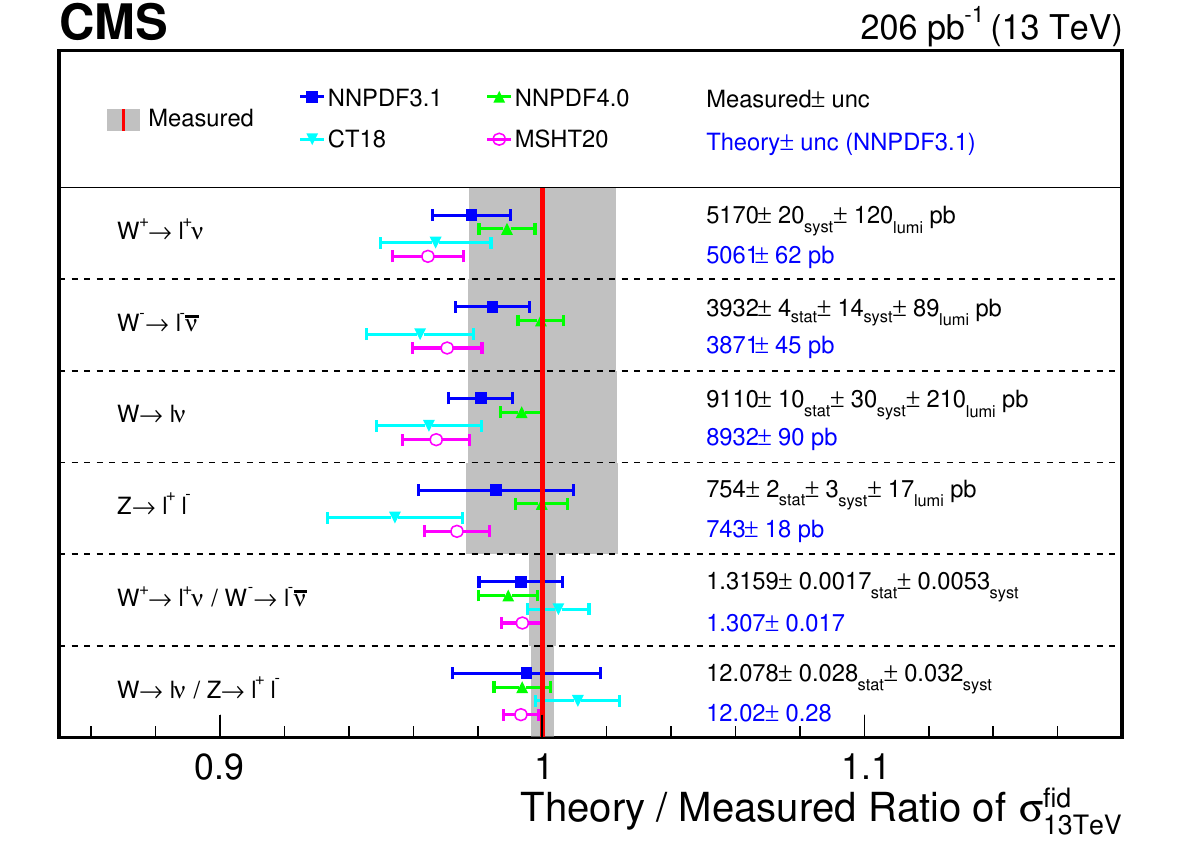}
    \caption{Comparisons of the fiducial inclusive cross sections and cross section ratios between measurements and the theoretical calculations from \textsc{DYturbo} with different PDF sets at 13\TeV. The gray band represents the total uncertainty of each measurement. The uncertainties in the theoretical predictions include the statistical uncertainty, and the PDF, \alpS, and renormalization and factorization scale uncertainties. The measured values and theoretical predictions (\textsc{DYturbo} with NNPDF 3.1 as the example) are also shown in the right part of the plot. The statistical uncertainty for $\PWp\to \Pell^{+}\PGn$ is negligible compared with the other uncertainties, therefore not included in this plot.}
    \label{fig:xsec_fid_13TeV}
\end{figure}

\begin{figure}[htbp]
    \centering
    \includegraphics[width=0.99\textwidth]{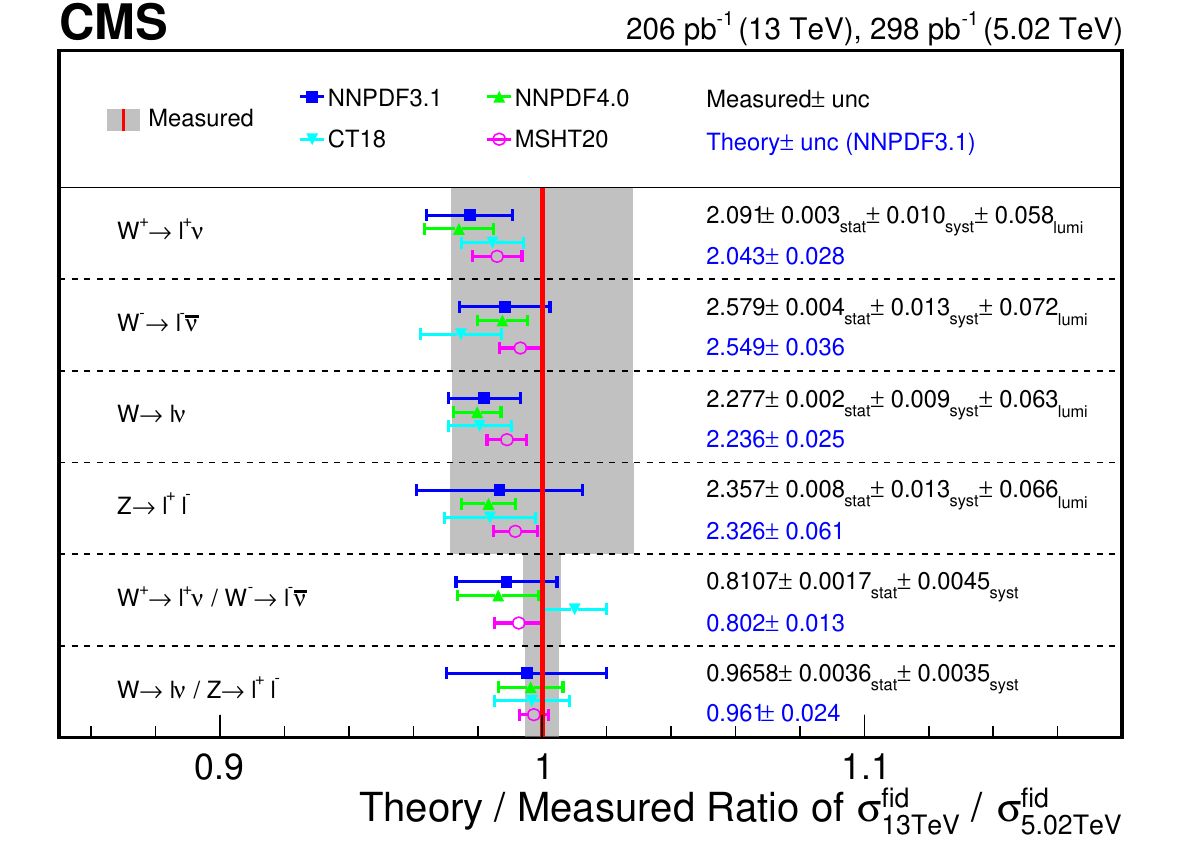}
    \caption{Comparisons of the fiducial inclusive cross section ratios and double ratios between 13 and 5.02\TeV, between measurements and the theoretical calculations from \textsc{DYturbo} with different PDF sets. The gray band represents the total uncertainty of each measurement. The uncertainties in the theoretical predictions include the statistical uncertainty, and the PDF, \alpS, and renormalization and factorization scale uncertainties. The measured values and theoretical predictions (\textsc{DYturbo} with NNPDF 3.1 as the example) are also shown in the right part of the plot.}
    \label{fig:xsec_fid_sqrtS}
\end{figure}

The results are presented as the measurements of the products of fiducial and total inclusive cross sections and the leptonic decay branching fractions. For fiducial measurements, the systematic uncertainties are reduced because no theoretical extrapolation of the result to the full phase space is performed. In measurements of the ratios of cross sections, some systematic uncertainties cancel, including the largest source in this measurement, the uncertainty in the integrated luminosity. A summary of measured and predicted \PWp, \PWm, \PW, and \PZ boson production fiducial inclusive cross sections times branching fractions, and \PWp to \PWm and \PW to \PZ ratios at 5.02 and 13\TeV are shown in Tables~\ref{tab:xsec_fid_5TeV} and~\ref{tab:xsec_fid_13TeV}, respectively. The ratios of these measurements between 13 and 5.02\TeV are shown in Table~\ref{tab:xsec_fid_sqrtS}. The ratios of theoretical predictions with different PDF sets to the measured fiducial inclusive cross section results are displayed in Fig.~\ref{fig:xsec_fid_5TeV} and~\ref{fig:xsec_fid_13TeV} at 5.02 and 13\TeV, respectively. The comparisons of ratios and double ratios between 13 and 5.02\TeV are provided in Fig.~\ref{fig:xsec_fid_sqrtS}.

It can be seen that the $\PWp/\PWm$ and $\PW/\PZ$ ratios agree well between measurements and theoretical calculations at 5.02, 13, and also between 13 and 5.02\TeV. For the absolute fiducial inclusive cross sections, the measurements agree well within uncertainties with theoretical cross sections at 5.02\TeV, shown in Fig.~\ref{fig:xsec_fid_5TeV}. At 13\TeV, shown in Fig.~\ref{fig:xsec_fid_13TeV}, the measured cross sections are larger than the predictions, among which NNPDF 4.0~\cite{NNPDF:2021njg} at NNLO in QCD provides the closest values. This also causes about the same level of theoretical underprediction of the ratios of \PW and \PZ boson production cross sections between 13 and 5.02\TeV, shown in Fig.~\ref{fig:xsec_fid_sqrtS}.

\begin{figure}[htbp]
    \centering
    \includegraphics[width=0.99\textwidth]{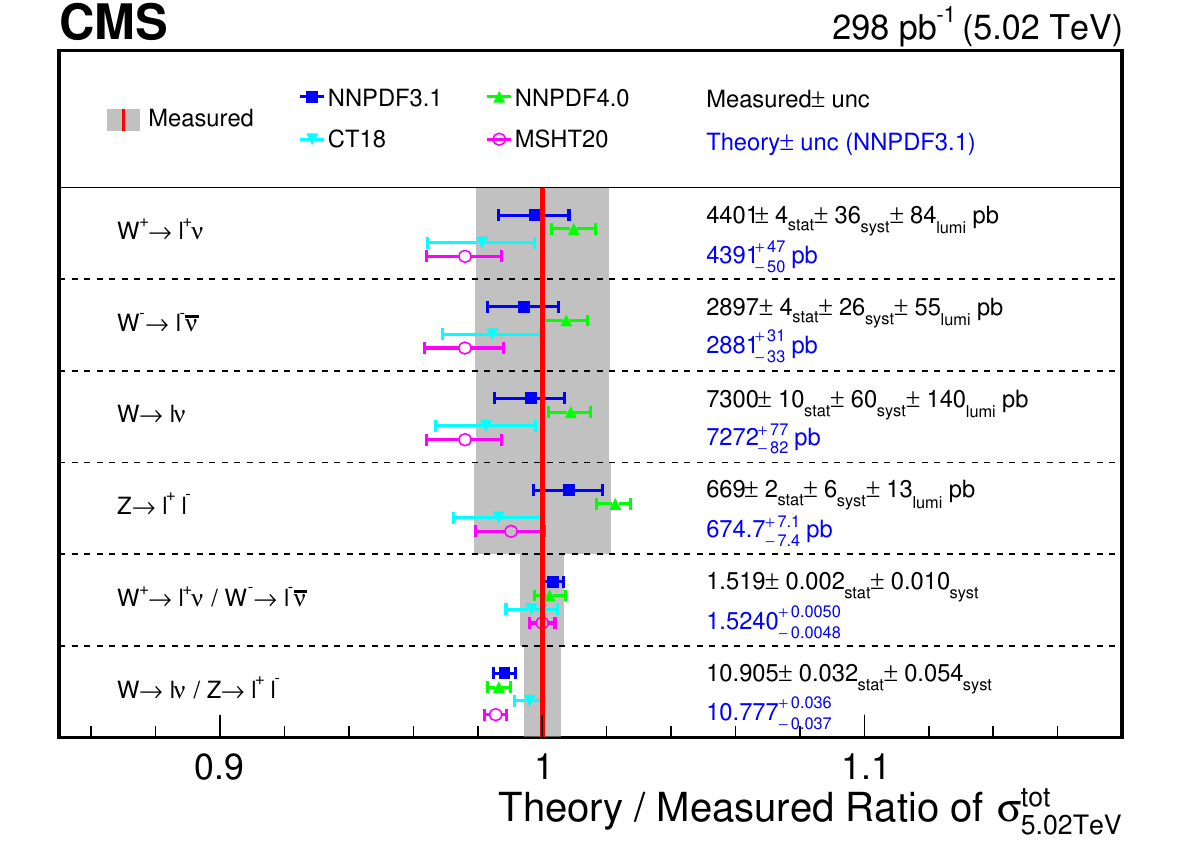}
    \caption{Comparisons of the total inclusive cross sections and cross section ratios between measurements and the theoretical calculations from \textsc{DYturbo} with different PDF sets at 5.02\TeV. The gray band represents the total uncertainty of each measurement. The uncertainties in the theoretical predictions include the statistical uncertainty, and the PDF, \alpS, and renormalization and factorization scale uncertainties. The measured values and theoretical predictions (\textsc{DYturbo} with NNPDF 3.1 as the example) are also shown in the right part of the plot.}
    \label{fig:xsec_inc_5TeV}
\end{figure}

\begin{figure}[htbp]
    \centering
    \includegraphics[width=0.99\textwidth]{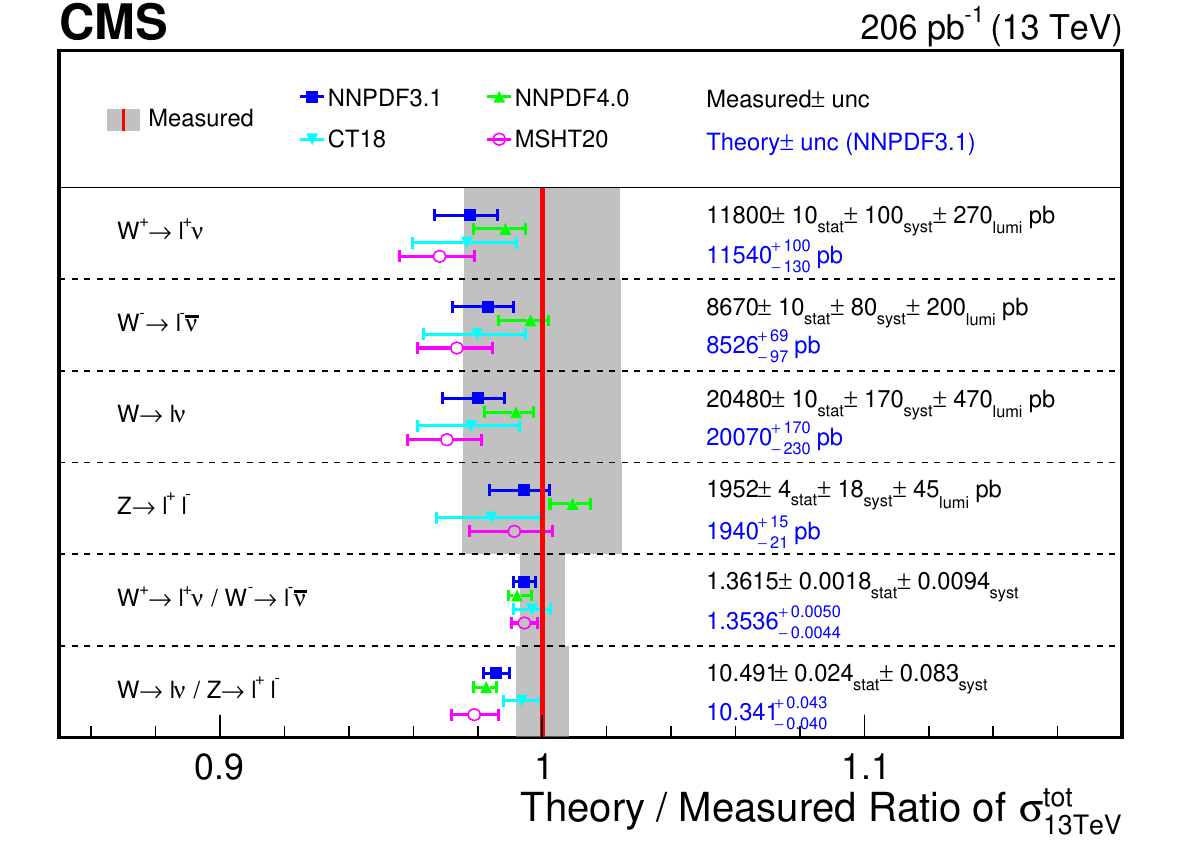}
    \caption{Comparisons of the total inclusive cross sections and cross section ratios between measurements and the theoretical calculations from \textsc{DYturbo} with different PDF sets at 13\TeV. The gray band represents the total uncertainty of each measurement. The uncertainties in the theoretical predictions include the statistical uncertainty, and the PDF, \alpS, and renormalization and factorization scale uncertainties. The measured values and theoretical predictions (\textsc{DYturbo} with NNPDF 3.1 as the example) are also shown in the right part of the plot.}
    \label{fig:xsec_inc_13TeV}
\end{figure}

\begin{figure}[htbp]
    \centering
    \includegraphics[width=0.99\textwidth]{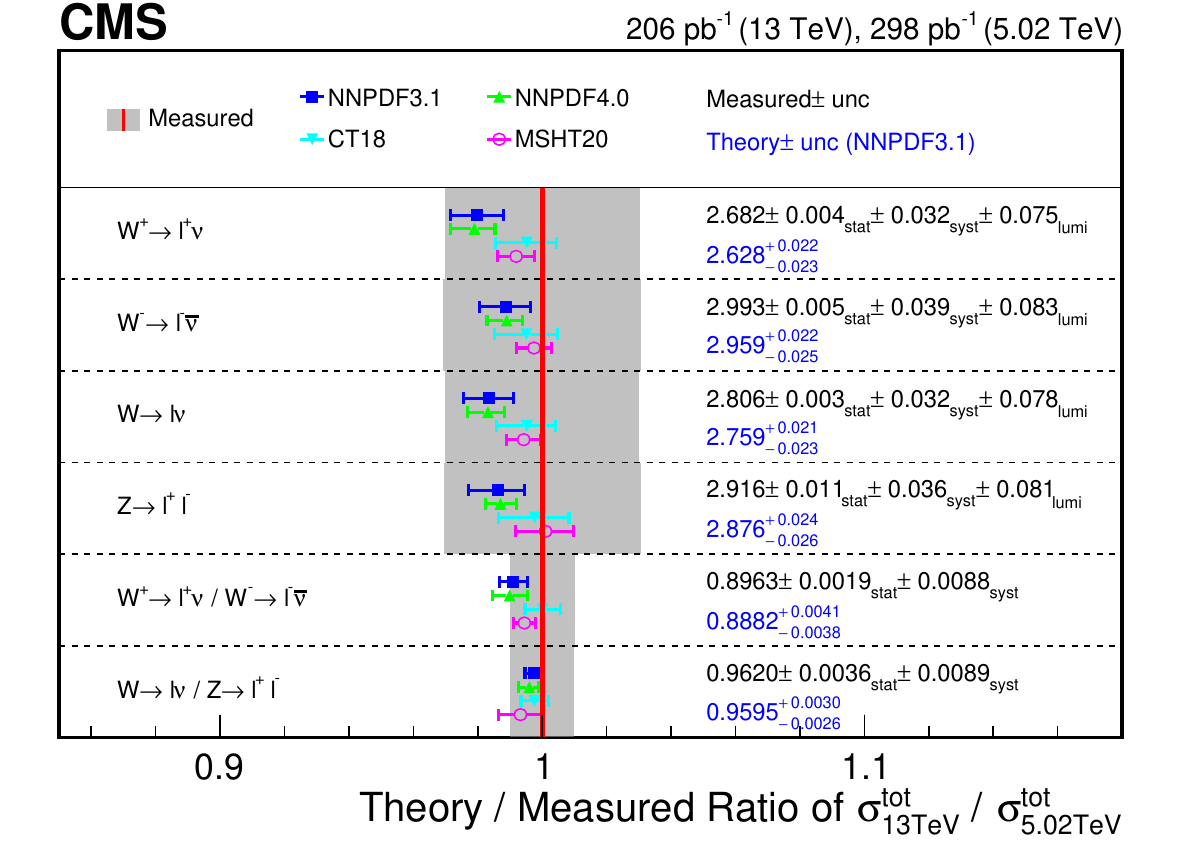}
    \caption{Comparisons of the total inclusive cross section ratios and double ratios between 13 and 5.02\TeV, between measurements and the theoretical calculations from \textsc{DYturbo} with different PDF sets. The gray band represents the total uncertainty of each measurement. The uncertainties in the theoretical predictions include the statistical uncertainty, and the PDF, \alpS, and renormalization and factorization scale uncertainties. The measured values and theoretical predictions (\textsc{DYturbo} with NNPDF 3.1 as the example) are also shown in the right part of the plot.}
    \label{fig:xsec_inc_sqrtS}
\end{figure}

Similarly to the fiducial inclusive cross sections, a summary of measured and predicted total inclusive cross sections for \PWp, \PWm, \PW, and \PZ boson production times branching fractions, \PW to \PZ and \PWp to \PWm ratios, and the cross section ratios and double ratios between 13 and 5.02\TeV are shown in Tables~\ref{tab:xsec_inc_5TeV},~\ref{tab:xsec_inc_13TeV} and~\ref{tab:xsec_inc_sqrtS}. The statistical uncertainties of theoretical predictions are largely reduced compared with fiducial inclusive cross section predictions, because of the switch from the Vegas algorithm~\cite{Lepage:1977sw} to interpolating functions for the numerical integration. The ratios of theoretical predictions with different PDF sets to the measured total inclusive cross section results are displayed in Fig.~\ref{fig:xsec_inc_5TeV} and~\ref{fig:xsec_inc_13TeV} at 5.02 and 13\TeV, respectively. The comparisons of ratios and double ratios between 13 and 5.02\TeV are provided in Fig.~\ref{fig:xsec_inc_sqrtS}. The results are also compared with theoretical predictions at $\mathrm{N}^{3}\mathrm{LO}$ in QCD~\cite{Baglio:2022wzu} and are consistent within uncertainties.

\begin{figure}[htbp]
    \centering
    \includegraphics[width=0.99\textwidth]{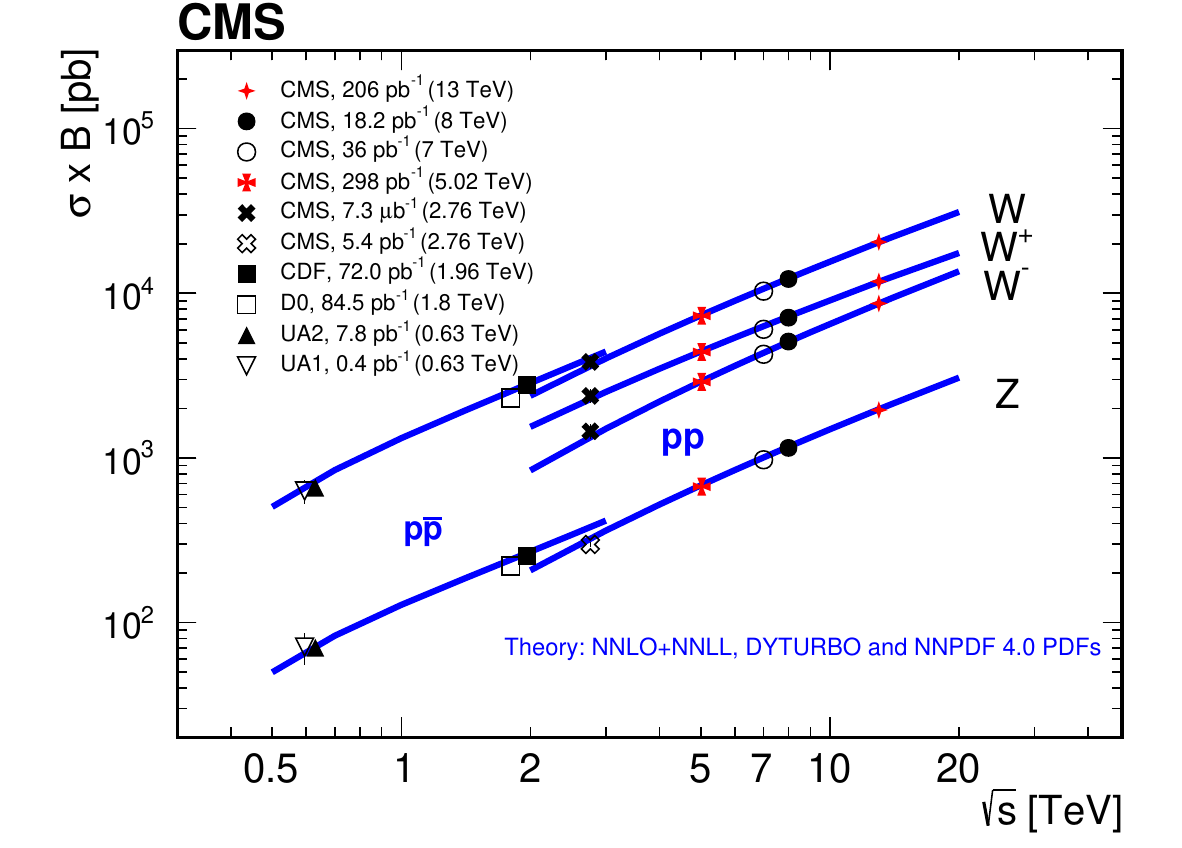}
    \caption{Summary of the measurements of the total inclusive cross sections for \PWp, \PWm, \PW, and \PZ boson production times branching fractions versus center-of-mass energy from CMS and experiments at lower-energy colliders. The vertical error bar represents the total uncertainty of each measurement.}
    \label{fig:xsec_vs_sqrtS}
\end{figure}

Figure~\ref{fig:xsec_vs_sqrtS} provides a summary of the measurements of the total inclusive cross sections for \PWp, \PWm, \PW, and \PZ boson production times branching fractions versus center-of-mass energy from CMS and experiments at lower-energy colliders~\cite{UA1:1987dfv,UA2:1990mxf,D0:1999qdf,CDF:2004sns,CMS:2012fgk,Chatrchyan_2015,CMS:2011aa,Chatrchyan:2014mua}. The NNPDF 4.0 is chosen as the reference prediction for the comparison as it provides the smallest uncertainties. The measurements are consistent with the theoretical predictions within uncertainties across different energy scales and different collisions (pp and p$\bar{\mathrm{p}}$).

\section{Summary}
\label{sec:summary}
Fiducial and total inclusive cross sections for \PW and \PZ boson production measured in proton-proton collisions at 5.02\TeV and 13\TeV are presented. Electron and muon decay modes are studied in the data collected with the CMS detector in 2017, in dedicated runs with reduced instantaneous luminosity. The data sets correspond to integrated luminosities of $298\pm6\pbinv$ at 5.02\TeV and $206\pm 5\pbinv$ at 13\TeV. Measured values of the products of the total inclusive cross sections and the branching fractions are $\sigma(\Pp\Pp \to \PW + \PX)\mathcal{B}(\PW \to \Pell\PGn ) = 7300 \pm 10\stat\pm 60\syst\pm 140\lum\pb$, and $\sigma(\Pp\Pp \to \PZ + \PX)\mathcal{B}(\PZ \to \Pell^+\Pell^-) = 669 \pm 2\stat\pm 6\syst \pm 13\lum\pb$ for the dilepton invariant mass in the range of 60--120\GeV at 5.02\TeV, and correspondingly $20480 \pm 10\stat\pm 170\syst\pm 470\lum\pb$ and $1952 \pm 4\stat\pm 18\syst\pm 45\lum\pb$ at 13\TeV. The measured values agree with cross section calculations at next-to-next-to-leading-order in perturbative quantum chromodynamics. Fiducial and total inclusive cross sections, ratios of cross sections of \PWp to \PWm, \PW to \PZ, and 13\TeV to 5.02\TeV measurements are reported. The fiducial inclusive cross sections for \PW and \PZ boson productions at 5.02\TeV achieve a precision of less than 2\%, which is the most precisely measured cross sections from the CMS experiment.

\begin{table}[htbp]
    \centering
    \caption{Comparison of the theoretical calculations and the measured fiducial inclusive cross sections and ratios at 5.02\TeV. The unit for cross sections is always \pb. The uncertainties in the theoretical predictions include the statistical uncertainty, and the PDF, \alpS, and renormalization and factorization scale uncertainties.}
    \label{tab:xsec_fid_5TeV}
    \cmsTable{
        \begin{tabular}{llllll}
            \hline
            {}                                  & Data                                                                & NNPDF3.1        & NNPDF4.0         & CT18            & MSHT20            \\
            \hline
            $\PWp\to \Pell^{+}\PGn$  & $2475\pm2_\text{stat}\pm8_\text{syst}\pm47_\text{lumi}$       & $2476\pm24$     & $2513\pm18$      & $2431\pm41$     & $2421\pm28$       \\
            $\PWm\to \Pell^{-}\PAGn$ & $1525\pm2_\text{stat}\pm5_\text{syst}\pm29_\text{lumi}$       & $1519\pm18$     & $1543.2\pm8.5$   & $1505\pm25$     & $1490\pm18$       \\
            $\PW\to \Pell\PGn$          & $4000\pm3_\text{stat}\pm11_\text{syst}\pm76_\text{lumi}$      & $3995\pm38$     & $4056\pm22$      & $3936\pm64$     & $3911\pm46$       \\
            $\PZ\to \Pell^{+}\Pell^{-}$ & $319.8\pm0.9_\text{stat}\pm1.2_\text{syst}\pm6.2_\text{lumi}$ & $319.5\pm3.7$   & $325.2\pm1.8$    & $310.2\pm4.9$   & $314.0\pm3.5$     \\
            $\PWp/\PWm$                   & $1.6232\pm0.0026_\text{stat}\pm0.0065_\text{syst}$              & $1.631\pm0.016$ & $1.628\pm0.013$  & $1.615\pm0.014$ & $1.6251\pm0.0076$ \\
            $\PW/\PZ$                           & $12.505\pm0.037_\text{stat}\pm0.032_\text{syst}$                & $12.51\pm0.12$  & $12.472\pm0.077$ & $12.69\pm0.13$  & $12.455\pm0.052$  \\
            \hline
        \end{tabular}
    }
\end{table}

\begin{table}[htbp]
    \centering
    \caption{Comparison of the theoretical calculations and the measured fiducial inclusive cross sections and ratios at 13\TeV. The unit for cross sections is always \pb. The uncertainties in the theoretical predictions include the statistical uncertainty, and the PDF, \alpS, and renormalization and factorization scale uncertainties. The statistical uncertainty for $\PWp\to \Pell^{+}\PGn$ is negligible compared with the other uncertainties.}
    \label{tab:xsec_fid_13TeV}
    \cmsTable{
        \begin{tabular}{llllll}
            \hline
            {}                                  & Data                                                             & NNPDF3.1        & NNPDF4.0        & CT18            & MSHT20            \\
            \hline
            $\PWp\to \Pell^{+}\PGn$  & $5170\pm20_\text{syst}\pm120_\text{lumi}$                    & $5061\pm62$     & $5118\pm45$     & $5003\pm89$     & $4991\pm57$       \\
            $\PWm\to \Pell^{-}\PAGn$ & $3932\pm4_\text{stat}\pm14_\text{syst}\pm89_\text{lumi}$   & $3871\pm45$     & $3930\pm28$     & $3783\pm65$     & $3816\pm43$       \\
            $\PW\to \Pell\PGn$          & $9110\pm10_\text{stat}\pm30_\text{syst}\pm210_\text{lumi}$ & $8932\pm90$     & $9048\pm61$     & $8790\pm150$    & $8807\pm95$       \\
            $\PZ\to \Pell^{+}\Pell^{-}$ & $754\pm2_\text{stat}\pm3_\text{syst}\pm17_\text{lumi}$     & $743\pm18$      & $753.9\pm6.0$   & $719\pm16$      & $734.0\pm7.7$     \\
            $\PWp/\PWm$                   & $1.3159\pm0.0017_\text{stat}\pm0.0053_\text{syst}$           & $1.307\pm0.017$ & $1.302\pm0.012$ & $1.322\pm0.013$ & $1.3078\pm0.0085$ \\
            $\PW/\PZ$                           & $12.078\pm0.028_\text{stat}\pm0.032_\text{syst}$             & $12.02\pm0.28$  & $12.00\pm0.11$  & $12.21\pm0.16$  & $11.998\pm0.065$  \\
            \hline
        \end{tabular}
    }
\end{table}

\begin{table}[htbp]
    \centering
    \caption{Comparison of the theoretical calculations and the measured fiducial inclusive cross section ratios between 13 and 5.02\TeV. The uncertainties in the theoretical predictions include the statistical uncertainty, and the PDF, \alpS, and renormalization and factorization scale uncertainties.}
    \label{tab:xsec_fid_sqrtS}
    \cmsTable{
        \begin{tabular}{llllll}
            \hline
            {}                                  & Data                                                                      & NNPDF3.1        & NNPDF4.0          & CT18              & MSHT20            \\
            \hline
            $\PWp\to \Pell^{+}\PGn$  & $2.091\pm0.003_\text{stat}\pm0.010_\text{syst}\pm0.058_\text{lumi}$ & $2.043\pm0.028$ & $2.037\pm0.022$   & $2.058\pm0.020$   & $2.061\pm0.016$   \\
            $\PWm\to \Pell^{-}\PAGn$ & $2.579\pm0.004_\text{stat}\pm0.013_\text{syst}\pm0.072_\text{lumi}$ & $2.549\pm0.036$ & $2.547\pm0.020$   & $2.514\pm0.032$   & $2.561\pm0.017$   \\
            $\PW\to \Pell\PGn$          & $2.277\pm0.002_\text{stat}\pm0.009_\text{syst}\pm0.063_\text{lumi}$ & $2.236\pm0.025$ & $2.231\pm0.017$   & $2.232\pm0.022$   & $2.252\pm0.014$   \\
            $\PZ\to \Pell^{+}\Pell^{-}$ & $2.357\pm0.008_\text{stat}\pm0.013_\text{syst}\pm0.066_\text{lumi}$ & $2.326\pm0.061$ & $2.318\pm0.020$   & $2.319\pm0.033$   & $2.338\pm0.016$   \\
            $\PWp/\PWm$                   & $0.8107\pm0.0017_\text{stat}\pm0.0045_\text{syst}$                    & $0.802\pm0.013$ & $0.800\pm0.010$   & $0.8188\pm0.0080$ & $0.8048\pm0.0061$ \\
            $\PW/\PZ$                           & $0.9658\pm0.0036_\text{stat}\pm0.0035_\text{syst}$                    & $0.961\pm0.024$ & $0.9623\pm0.0097$ & $0.963\pm0.011$   & $0.9633\pm0.0044$ \\
            \hline
        \end{tabular}
    }
\end{table}

\begin{table}[htbp]
    \centering
    \caption{Comparison of the theoretical calculations and the measured total inclusive cross sections and ratios at 5.02\TeV. The unit for cross sections is always \pb. The uncertainties in the theoretical predictions include the statistical uncertainty, and the PDF, \alpS, and renormalization and factorization scale uncertainties.}
    \label{tab:xsec_inc_5TeV}
    \cmsTable{
        \begin{tabular}{llllll}
            \hline
            {}                                  & Data                                                             & NNPDF3.1                     & NNPDF4.0                     & CT18                       & MSHT20                     \\
            \hline
            $\PWp\to \Pell^{+}\PGn$  & $4401\pm4_\text{stat}\pm36_\text{syst}\pm84_\text{lumi}$   & $4391^{+47}_{-50}$           & $4444^{+29}_{-30}$           & $4319^{+72}_{-74}$         & $4295^{+50}_{-53}$         \\
            $\PWm\to \Pell^{-}\PAGn$ & $2897\pm4_\text{stat}\pm26_\text{syst}\pm55_\text{lumi}$   & $2881^{+31}_{-33}$           & $2919^{+19}_{-22}$           & $2853^{+43}_{-45}$         & $2828^{+35}_{-37}$         \\
            $\PW\to \Pell\PGn$          & $7300\pm10_\text{stat}\pm60_\text{syst}\pm140_\text{lumi}$ & $7272^{+77}_{-82}$           & $7363^{+45}_{-50}$           & $7170^{+110}_{-120}$       & $7123^{+84}_{-88}$         \\
            $\PZ\to \Pell^{+}\Pell^{-}$ & $669\pm2_\text{stat}\pm6_\text{syst}\pm13_\text{lumi}$     & $674.7^{+7.1}_{-7.4}$        & $684.4^{+3.2}_{-3.8}$        & $660.2^{+9.2}_{-9.5}$      & $662.7^{+7.0}_{-7.3}$      \\
            $\PWp/\PWm$                   & $1.519\pm0.002_\text{stat}\pm0.010_\text{syst}$              & $1.5240^{+0.0050}_{-0.0048}$ & $1.5225^{+0.0076}_{-0.0074}$ & $1.514\pm0.012$            & $1.5190\pm0.0060$          \\
            $\PW/\PZ$                           & $10.905\pm0.032_\text{stat}\pm0.054_\text{syst}$             & $10.777^{+0.036}_{-0.037}$   & $10.758^{+0.039}_{-0.037}$   & $10.862^{+0.050}_{-0.051}$ & $10.748^{+0.036}_{-0.037}$ \\
            \hline
        \end{tabular}
    }
\end{table}

\begin{table}[htbp]
    \centering
    \caption{Comparison of the theoretical calculations and the measured total inclusive cross sections and ratios at 13\TeV. The unit for cross sections is always \pb. The uncertainties in the theoretical predictions include the statistical uncertainty, and the PDF, \alpS, and renormalization and factorization scale uncertainties.}
    \label{tab:xsec_inc_13TeV}
    \cmsTable{
        \begin{tabular}{llllll}
            \hline
            {}                                  & Data                                                               & NNPDF3.1                     & NNPDF4.0                     & CT18                       & MSHT20                     \\
            \hline
            $\PWp\to \Pell^{+}\PGn$  & $11800\pm10_\text{stat}\pm100_\text{syst}\pm270_\text{lumi}$ & $11540^{+100}_{-130}$        & $11670^{+70}_{-120}$         & $11530^{+180}_{-200}$      & $11430^{+130}_{-150}$      \\
            $\PWm\to \Pell^{-}\PAGn$ & $8670\pm10_\text{stat}\pm80_\text{syst}\pm200_\text{lumi}$   & $8526^{+69}_{-97}$           & $8639^{+49}_{-85}$           & $8490^{+130}_{-140}$       & $8440^{+100}_{-110}$       \\
            $\PW\to \Pell\PGn$          & $20480\pm10_\text{stat}\pm170_\text{syst}\pm470_\text{lumi}$ & $20070^{+170}_{-230}$        & $20310^{+110}_{-200}$        & $20020^{+310}_{-340}$      & $19870^{+220}_{-250}$      \\
            $\PZ\to \Pell^{+}\Pell^{-}$ & $1952\pm4_\text{stat}\pm18_\text{syst}\pm45_\text{lumi}$     & $1940^{+15}_{-21}$           & $1970^{+11}_{-14}$           & $1921^{+30}_{-33}$         & $1935^{+23}_{-27}$         \\
            $\PWp/\PWm$                   & $1.3615\pm0.0018_\text{stat}\pm0.0094_\text{syst}$             & $1.3536^{+0.0050}_{-0.0044}$ & $1.3508^{+0.0062}_{-0.0038}$ & $1.3571\pm0.0077$          & $1.3539\pm0.0056$          \\
            $\PW/\PZ$                           & $10.491\pm0.024_\text{stat}\pm0.083_\text{syst}$               & $10.341^{+0.043}_{-0.040}$   & $10.309^{+0.033}_{-0.042}$   & $10.424^{+0.062}_{-0.060}$ & $10.270^{+0.078}_{-0.074}$ \\
            \hline
        \end{tabular}
    }
\end{table}

\begin{table}[htbp]
    \centering
    \caption{Comparison of the theoretical calculations and the measured total inclusive cross section ratios and double ratios between 13 and 5.02\TeV. The uncertainties in the theoretical predictions include the statistical uncertainty, and the PDF, \alpS, and renormalization and factorization scale uncertainties.}
    \label{tab:xsec_inc_sqrtS}
    \cmsTable{
        \begin{tabular}{llllll}
            \hline
            {}                                  & Data                                                                      & NNPDF3.1                     & NNPDF4.0                     & CT18                         & MSHT20                       \\
            \hline
            $\PWp\to \Pell^{+}\PGn$  & $2.682\pm0.004_\text{stat}\pm0.032_\text{syst}\pm0.075_\text{lumi}$ & $2.628^{+0.022}_{-0.023}$    & $2.626^{+0.017}_{-0.020}$    & $2.669^{+0.025}_{-0.026}$    & $2.661^{+0.015}_{-0.016}$    \\
            $\PWm\to \Pell^{-}\PAGn$ & $2.993\pm0.005_\text{stat}\pm0.039_\text{syst}\pm0.083_\text{lumi}$ & $2.959^{+0.022}_{-0.025}$    & $2.960^{+0.014}_{-0.018}$    & $2.978^{+0.029}_{-0.030}$    & $2.985\pm0.016$              \\
            $\PW\to \Pell\PGn$          & $2.806\pm0.003_\text{stat}\pm0.032_\text{syst}\pm0.078_\text{lumi}$ & $2.759^{+0.021}_{-0.023}$    & $2.758^{+0.014}_{-0.018}$    & $2.792^{+0.025}_{-0.026}$    & $2.789\pm0.015$              \\
            $\PZ\to \Pell^{+}\Pell^{-}$ & $2.916\pm0.011_\text{stat}\pm0.036_\text{syst}\pm0.081_\text{lumi}$ & $2.876^{+0.024}_{-0.026}$    & $2.879\pm0.014$              & $2.910^{+0.031}_{-0.033}$    & $2.919^{+0.025}_{-0.027}$    \\
            $\PWp/\PWm$                   & $0.8963\pm0.0019_\text{stat}\pm0.0088_\text{syst}$                    & $0.8882^{+0.0041}_{-0.0038}$ & $0.8872^{+0.0051}_{-0.0047}$ & $0.8964\pm0.0049$            & $0.8913\pm0.0030$            \\
            $\PW/\PZ$                           & $0.9620\pm0.0036_\text{stat}\pm0.0089_\text{syst}$                    & $0.9595^{+0.0030}_{-0.0026}$ & $0.9582^{+0.0026}_{-0.0032}$ & $0.9597^{+0.0041}_{-0.0040}$ & $0.9555^{+0.0068}_{-0.0066}$ \\
            \hline
        \end{tabular}
    }
\end{table}

\begin{acknowledgments}
We congratulate our colleagues in the CERN accelerator departments for the excellent performance of the LHC and thank the technical and administrative staffs at CERN and at other CMS institutes for their contributions to the success of the CMS effort. In addition, we gratefully acknowledge the computing centers and personnel of the Worldwide LHC Computing Grid and other centers for delivering so effectively the computing infrastructure essential to our analyses. Finally, we acknowledge the enduring support for the construction and operation of the LHC, the CMS detector, and the supporting computing infrastructure provided by the following funding agencies: SC (Armenia), BMBWF and FWF (Austria); FNRS and FWO (Belgium); CNPq, CAPES, FAPERJ, FAPERGS, and FAPESP (Brazil); MES and BNSF (Bulgaria); CERN; CAS, MoST, and NSFC (China); MINCIENCIAS (Colombia); MSES and CSF (Croatia); RIF (Cyprus); SENESCYT (Ecuador); ERC PRG, RVTT3 and MoER TK202 (Estonia); Academy of Finland, MEC, and HIP (Finland); CEA and CNRS/IN2P3 (France); SRNSF (Georgia); BMBF, DFG, and HGF (Germany); GSRI (Greece); NKFIH (Hungary); DAE and DST (India); IPM (Iran); SFI (Ireland); INFN (Italy); MSIP and NRF (Republic of Korea); MES (Latvia); LMTLT (Lithuania); MOE and UM (Malaysia); BUAP, CINVESTAV, CONACYT, LNS, SEP, and UASLP-FAI (Mexico); MOS (Montenegro); MBIE (New Zealand); PAEC (Pakistan); MES and NSC (Poland); FCT (Portugal); MESTD (Serbia); MCIN/AEI and PCTI (Spain); MOSTR (Sri Lanka); Swiss Funding Agencies (Switzerland); MST (Taipei); MHESI and NSTDA (Thailand); TUBITAK and TENMAK (Turkey); NASU (Ukraine); STFC (United Kingdom); DOE and NSF (USA).

\hyphenation{Rachada-pisek} Individuals have received support from the Marie-Curie program and the European Research Council and Horizon 2020 Grant, contract Nos.\ 675440, 724704, 752730, 758316, 765710, 824093, 101115353, 101002207, and COST Action CA16108 (European Union); the Leventis Foundation; the Alfred P.\ Sloan Foundation; the Alexander von Humboldt Foundation; the Science Committee, project no. 22rl-037 (Armenia); the Belgian Federal Science Policy Office; the Fonds pour la Formation \`a la Recherche dans l'Industrie et dans l'Agriculture (FRIA-Belgium); the F.R.S.-FNRS and FWO (Belgium) under the ``Excellence of Science -- EOS" -- be.h project n.\ 30820817; the Beijing Municipal Science \& Technology Commission, No. Z191100007219010 and Fundamental Research Funds for the Central Universities (China); the Ministry of Education, Youth and Sports (MEYS) of the Czech Republic; the Shota Rustaveli National Science Foundation, grant FR-22-985 (Georgia); the Deutsche Forschungsgemeinschaft (DFG), among others, under Germany's Excellence Strategy -- EXC 2121 ``Quantum Universe" -- 390833306, and under project number 400140256 - GRK2497; the Hellenic Foundation for Research and Innovation (HFRI), Project Number 2288 (Greece); the Hungarian Academy of Sciences, the New National Excellence Program - \'UNKP, the NKFIH research grants K 131991, K 133046, K 138136, K 143460, K 143477, K 146913, K 146914, K 147048, 2020-2.2.1-ED-2021-00181, and TKP2021-NKTA-64 (Hungary); the Council of Science and Industrial Research, India; ICSC -- National Research Center for High Performance Computing, Big Data and Quantum Computing and FAIR -- Future Artificial Intelligence Research, funded by the NextGenerationEU program (Italy); the Latvian Council of Science; the Ministry of Education and Science, project no. 2022/WK/14, and the National Science Center, contracts Opus 2021/41/B/ST2/01369 and 2021/43/B/ST2/01552 (Poland); the Funda\c{c}\~ao para a Ci\^encia e a Tecnologia, grant CEECIND/01334/2018 (Portugal); the National Priorities Research Program by Qatar National Research Fund; MCIN/AEI/10.13039/501100011033, ERDF ``a way of making Europe", and the Programa Estatal de Fomento de la Investigaci{\'o}n Cient{\'i}fica y T{\'e}cnica de Excelencia Mar\'{\i}a de Maeztu, grant MDM-2017-0765 and Programa Severo Ochoa del Principado de Asturias (Spain); the Chulalongkorn Academic into Its 2nd Century Project Advancement Project, and the National Science, Research and Innovation Fund via the Program Management Unit for Human Resources \& Institutional Development, Research and Innovation, grant B39G670016 (Thailand); the Kavli Foundation; the Nvidia Corporation; the SuperMicro Corporation; the Welch Foundation, contract C-1845; and the Weston Havens Foundation (USA).
\end{acknowledgments}

\bibliography{auto_generated}
\cleardoublepage \appendix\section{The CMS Collaboration \label{app:collab}}\begin{sloppypar}\hyphenpenalty=5000\widowpenalty=500\clubpenalty=5000\input{SMP-20-004-public-authorlist.tex}\end{sloppypar}
\end{document}

%% file: SMP-20-004-public-authorlist.tex
\cmsinstitute{Yerevan Physics Institute, Yerevan, Armenia}
{\tolerance=6000
A.~Hayrapetyan, A.~Tumasyan\cmsAuthorMark{1}\cmsorcid{0009-0000-0684-6742}
\par}
\cmsinstitute{Institut f\"{u}r Hochenergiephysik, Vienna, Austria}
{\tolerance=6000
W.~Adam\cmsorcid{0000-0001-9099-4341}, J.W.~Andrejkovic, T.~Bergauer\cmsorcid{0000-0002-5786-0293}, S.~Chatterjee\cmsorcid{0000-0003-2660-0349}, K.~Damanakis\cmsorcid{0000-0001-5389-2872}, M.~Dragicevic\cmsorcid{0000-0003-1967-6783}, P.S.~Hussain\cmsorcid{0000-0002-4825-5278}, M.~Jeitler\cmsAuthorMark{2}\cmsorcid{0000-0002-5141-9560}, N.~Krammer\cmsorcid{0000-0002-0548-0985}, A.~Li\cmsorcid{0000-0002-4547-116X}, D.~Liko\cmsorcid{0000-0002-3380-473X}, I.~Mikulec\cmsorcid{0000-0003-0385-2746}, J.~Schieck\cmsAuthorMark{2}\cmsorcid{0000-0002-1058-8093}, R.~Sch\"{o}fbeck\cmsorcid{0000-0002-2332-8784}, D.~Schwarz\cmsorcid{0000-0002-3821-7331}, M.~Sonawane\cmsorcid{0000-0003-0510-7010}, S.~Templ\cmsorcid{0000-0003-3137-5692}, W.~Waltenberger\cmsorcid{0000-0002-6215-7228}, C.-E.~Wulz\cmsAuthorMark{2}\cmsorcid{0000-0001-9226-5812}
% ...
\par}
\vskip\cmsinstskip
\dag:~Deceased\\
$^{1}$Also at Yerevan State University, Yerevan, Armenia\\
$^{2}$Also at TU Wien, Vienna, Austria\\
% ...


%% file: SMP-20-004_temp.bbl
\providecommand{\href}[2]{#2}\begingroup\raggedright\begin{thebibliography}{10}%
\makeatletter
\providecommand{\hrefCMSnoop }[0]{\@secondoftwo}%
\makeatother
\providecommand{\doi}{\texttt{doi:}\begingroup \urlstyle{tt}\Url}

\bibitem{Butterworth:2015oua}
\hrefCMSnoop {}{J.~Butterworth { et~al.}, ``{PDF4LHC recommendations for LHC
  Run II}'',} \textit{ J. Phys. G} \textbf{ 43} (2016) 023001,
  \href{http://dx.doi.org/10.1088/0954-3899/43/2/023001}{\doi{10.1088/0954-3899/43/2/023001}},
\href{http://www.arXiv.org/abs/1510.03865}{\texttt{arXiv:1510.03865}}.

\bibitem{Bailey:2020ooq}
S.~Bailey\hrefCMSnoop {}{ { et~al.}, ``{Parton distributions from LHC, HERA,
  Tevatron and fixed target data: MSHT20 PDFs}'',} \textit{ Eur. Phys. J. C}
  \textbf{ 81} (2021) 341,
  \href{http://dx.doi.org/10.1140/epjc/s10052-021-09057-0}{\doi{10.1140/epjc/s10052-021-09057-0}},
  \href{http://www.arXiv.org/abs/2012.04684}{\texttt{arXiv:2012.04684}}.

\bibitem{Hou:2019qau}
\hrefCMSnoop {}{T.-J. Hou { et~al.}, ``{Progress in the CTEQ-TEA NNLO global
  QCD analysis}'',} 2019.
  \href{http://www.arXiv.org/abs/1908.11394}{\texttt{arXiv:1908.11394}}.

\bibitem{NNPDF:2017mvq}
\hrefCMSnoop {}{{NNPDF} Collaboration, ``{Parton distributions from
  high-precision collider data}'',} \textit{ Eur. Phys. J. C} \textbf{ 77}
  (2017) 663,
  \href{http://dx.doi.org/10.1140/epjc/s10052-017-5199-5}{\doi{10.1140/epjc/s10052-017-5199-5}},
  \href{http://www.arXiv.org/abs/1706.00428}{\texttt{arXiv:1706.00428}}.

\bibitem{Melnikov:2006kv}
\hrefCMSnoop {}{K.~Melnikov and F.~Petriello, ``{Electroweak gauge boson
  production at hadron colliders through
  $\mathcal{O}(\alpha_{\mathrm{s}}^2)$}'',} \textit{ Phys. Rev. D} \textbf{ 74}
  (2006) 114017,
  \href{http://dx.doi.org/10.1103/PhysRevD.74.114017}{\doi{10.1103/PhysRevD.74.114017}},
\href{http://www.arXiv.org/abs/hep-ph/0609070}{\texttt{arXiv:hep-ph/0609070}}.

\bibitem{Catani:2009sm}
S.~Catani\hrefCMSnoop {}{ { et~al.}, ``{Vector boson production at hadron
  colliders: a fully exclusive QCD calculation at NNLO}'',} \textit{ Phys. Rev.
  Lett.} \textbf{ 103} (2009) 082001,
  \href{http://dx.doi.org/10.1103/PhysRevLett.103.082001}{\doi{10.1103/PhysRevLett.103.082001}},
\href{http://www.arXiv.org/abs/0903.2120}{\texttt{arXiv:0903.2120}}.

\bibitem{Baglio:2022wzu}
\hrefCMSnoop {}{J.~Baglio, C.~Duhr, B.~Mistlberger, and R.~Szafron,
  ``{Inclusive production cross sections at N$^{3}$LO}'',} \textit{ JHEP}
  \textbf{ 12} (2022) 066,
  \href{http://dx.doi.org/10.1007/JHEP12(2022)066}{\doi{10.1007/JHEP12(2022)066}},
  \href{http://www.arXiv.org/abs/2209.06138}{\texttt{arXiv:2209.06138}}.

\bibitem{Anastasiou:2003ds}
\hrefCMSnoop {}{C.~Anastasiou, L.~Dixon, K.~Melnikov, and F.~Petriello, ``{High
  precision QCD at hadron colliders: Electroweak gauge boson rapidity
  distributions at NNLO}'',} \textit{ Phys. Rev. D} \textbf{ 69} (2004) 094008,
  \href{http://dx.doi.org/10.1103/PhysRevD.69.094008}{\doi{10.1103/PhysRevD.69.094008}},
\href{http://www.arXiv.org/abs/hep-ph/0312266}{\texttt{arXiv:hep-ph/0312266}}.

\bibitem{Dittmaier:2014qza}
\hrefCMSnoop {}{S.~Dittmaier, A.~Huss, and C.~Schwinn, ``{Mixed QCD-electroweak
  $\mathcal{O}(\alpha_s\alpha)$ corrections to Drell-Yan processes in the
  resonance region: pole approximation and non-factorizable corrections}'',}
  \textit{ Nucl. Phys. B} \textbf{ 885} (2014) 318,
  \href{http://dx.doi.org/10.1016/j.nuclphysb.2014.05.027}{\doi{10.1016/j.nuclphysb.2014.05.027}},
\href{http://www.arXiv.org/abs/1403.3216}{\texttt{arXiv:1403.3216}}.

\bibitem{Lindert:2017olm}
\hrefCMSnoop {}{J.~M. Lindert { et~al.}, ``{Precise predictions for V+jets dark
  matter backgrounds}'',} \textit{ Eur. Phys. J. C} \textbf{ 77} (2017) 829,
  \href{http://dx.doi.org/10.1140/epjc/s10052-017-5389-1}{\doi{10.1140/epjc/s10052-017-5389-1}},
\href{http://www.arXiv.org/abs/1705.04664}{\texttt{arXiv:1705.04664}}.

\bibitem{Chatrchyan:2008zzk}
\hrefCMSnoop {}{{CMS Collaboration}, ``The {CMS} experiment at the {CERN}
  {LHC}'',} \textit{ JINST} \textbf{ 3} (2008) S08004,
  \href{http://dx.doi.org/10.1088/1748-0221/3/08/S08004}{\doi{10.1088/1748-0221/3/08/S08004}}.

\bibitem{Chatrchyan_2015}
\hrefCMSnoop {}{{CMS Collaboration}, ``{Study of Z production in PbPb and pp
  collisions at $ \sqrt{s_{\mathrm{NN}}}=2.76 $ TeV in the dimuon and
  dielectron decay channels}'',} \textit{ JHEP} \textbf{ 03} (2015) 022,
  \href{http://dx.doi.org/10.1007/jhep03(2015)022}{\doi{10.1007/jhep03(2015)022}},
  \href{http://www.arXiv.org/abs/1410.4825}{\texttt{arXiv:1410.4825}}.

\bibitem{CMS:2012fgk}
\hrefCMSnoop {}{{CMS Collaboration}, ``{Study of W boson production in PbPb and
  pp collisions at $\sqrt{s_{\mathrm{NN}}}=2.76$ TeV}'',} \textit{ Phys. Lett.
  B} \textbf{ 715} (2012) 66,
  \href{http://dx.doi.org/10.1016/j.physletb.2012.07.025}{\doi{10.1016/j.physletb.2012.07.025}},
  \href{http://www.arXiv.org/abs/1205.6334}{\texttt{arXiv:1205.6334}}.

\bibitem{Aad:2019bdc}
\hrefCMSnoop {}{{ATLAS Collaboration}, ``{Measurement of $\PW^{\pm}$-boson and
  $\PZ$-boson production cross-sections in $\Pp\Pp$ collisions at
  $\sqrt{s}=2.76$\TeV with the ATLAS detector}'',} \textit{ Eur. Phys. J. C}
  \textbf{ 79} (2019) 901,
  \href{http://dx.doi.org/10.1140/epjc/s10052-019-7399-7}{\doi{10.1140/epjc/s10052-019-7399-7}},
\href{http://www.arXiv.org/abs/1907.03567}{\texttt{arXiv:1907.03567}}.

\bibitem{Aaboud:2018nic}
\hrefCMSnoop {}{{ATLAS Collaboration}, ``{Measurements of $\PW$ and $\PZ$ boson
  production in $\Pp\Pp$ collisions at $\sqrt{s}=5.02$\TeV with the ATLAS
  detector}'',} \textit{ Eur. Phys. J. C} \textbf{ 79} (2019) 128,
  \href{http://dx.doi.org/10.1140/epjc/s10052-019-6622-x}{\doi{10.1140/epjc/s10052-019-6622-x}},
  \href{http://www.arXiv.org/abs/1810.08424}{\texttt{arXiv:1810.08424}}.
[Erratum: \DOI{10.1140/epjc/s10052-019-6870-9}].

\bibitem{Aaboud:2016btc}
\hrefCMSnoop {}{{ATLAS Collaboration}, ``{Precision measurement and
  interpretation of inclusive $\PW^+$ , $\PW^-$ and $\PZ/\gamma ^*$ production
  cross sections with the ATLAS detector}'',} \textit{ Eur. Phys. J. C}
  \textbf{ 77} (2017) 367,
  \href{http://dx.doi.org/10.1140/epjc/s10052-017-4911-9}{\doi{10.1140/epjc/s10052-017-4911-9}},
\href{http://www.arXiv.org/abs/1612.03016}{\texttt{arXiv:1612.03016}}.

\bibitem{CMS:2011aa}
\hrefCMSnoop {}{{CMS Collaboration}, ``{Measurement of the inclusive $\PW$ and
  $\PZ$ production cross sections in $\Pp\Pp$ collisions at
  $\sqrt{s}=7$\TeV}'',} \textit{ JHEP} \textbf{ 10} (2011) 132,
  \href{http://dx.doi.org/10.1007/JHEP10(2011)132}{\doi{10.1007/JHEP10(2011)132}},
\href{http://www.arXiv.org/abs/1107.4789}{\texttt{arXiv:1107.4789}}.

\bibitem{Aaij:2014wba}
\hrefCMSnoop {}{{LHCb Collaboration}, ``{Measurement of the forward $\PW$ boson
  cross-section in $\Pp\Pp$ collisions at $\sqrt{s} = 7$\TeV}'',} \textit{
  JHEP} \textbf{ 12} (2014) 079,
  \href{http://dx.doi.org/10.1007/JHEP12(2014)079}{\doi{10.1007/JHEP12(2014)079}},
\href{http://www.arXiv.org/abs/1408.4354}{\texttt{arXiv:1408.4354}}.

\bibitem{Aaij:2012mda}
\hrefCMSnoop {}{{LHCb Collaboration}, ``{Measurement of the cross-section for
  $\PZ \to e^+e^-$ production in $\Pp\Pp$ collisions at $\sqrt{s}=7$\TeV}'',}
  \textit{ JHEP} \textbf{ 02} (2013) 106,
  \href{http://dx.doi.org/10.1007/JHEP02(2013)106}{\doi{10.1007/JHEP02(2013)106}},
\href{http://www.arXiv.org/abs/1212.4620}{\texttt{arXiv:1212.4620}}.

\bibitem{Aaij:2015gna}
\hrefCMSnoop {}{{LHCb Collaboration}, ``{Measurement of the forward $\PZ$ boson
  production cross-section in $\Pp\Pp$ collisions at $\sqrt{s}=7$\TeV}'',}
  \textit{ JHEP} \textbf{ 08} (2015) 039,
  \href{http://dx.doi.org/10.1007/JHEP08(2015)039}{\doi{10.1007/JHEP08(2015)039}},
\href{http://www.arXiv.org/abs/1505.07024}{\texttt{arXiv:1505.07024}}.

\bibitem{Aaij:2016qqz}
\hrefCMSnoop {}{{LHCb Collaboration}, ``{Measurement of forward $\PW\to e\nu$
  production in $\Pp\Pp$ collisions at $\sqrt{s}=8$\TeV}'',} \textit{ JHEP}
  \textbf{ 10} (2016) 030,
  \href{http://dx.doi.org/10.1007/JHEP10(2016)030}{\doi{10.1007/JHEP10(2016)030}},
\href{http://www.arXiv.org/abs/1608.01484}{\texttt{arXiv:1608.01484}}.

\bibitem{Aaij:2016mgv}
\hrefCMSnoop {}{{LHCb Collaboration}, ``{Measurement of the forward $\PZ$ boson
  production cross-section in pp collisions at $\sqrt{s} = 13$\TeV}'',}
  \textit{ JHEP} \textbf{ 09} (2016) 136,
  \href{http://dx.doi.org/10.1007/JHEP09(2016)136}{\doi{10.1007/JHEP09(2016)136}},
\href{http://www.arXiv.org/abs/1607.06495}{\texttt{arXiv:1607.06495}}.

\bibitem{Aaij:2015zlq}
\hrefCMSnoop {}{{LHCb Collaboration}, ``{Measurement of forward $\PW$ and $\PZ$
  boson production in $\Pp\Pp$ collisions at $\sqrt{s}=8$\TeV}'',} \textit{
  JHEP} \textbf{ 01} (2016) 155,
  \href{http://dx.doi.org/10.1007/JHEP01(2016)155}{\doi{10.1007/JHEP01(2016)155}},
\href{http://www.arXiv.org/abs/1511.08039}{\texttt{arXiv:1511.08039}}.

\bibitem{Chatrchyan:2014mua}
\hrefCMSnoop {}{{CMS Collaboration}, ``{Measurement of inclusive $\PW$ and
  $\PZ$ boson production cross sections in pp collisions at $\sqrt{s}$ = 8
  TeV}'',} \textit{ Phys. Rev. Lett.} \textbf{ 112} (2014) 191802,
  \href{http://dx.doi.org/10.1103/PhysRevLett.112.191802}{\doi{10.1103/PhysRevLett.112.191802}},
\href{http://www.arXiv.org/abs/1402.0923}{\texttt{arXiv:1402.0923}}.

\bibitem{Aad:2016naf}
\hrefCMSnoop {}{{ATLAS Collaboration}, ``{Measurement of $\PW^{\pm}$ and
  $\PZ$-boson production cross sections in $\Pp\Pp$ collisions at
  $\sqrt{s}=13$\TeV with the ATLAS detector}'',} \textit{ Phys. Lett. B}
  \textbf{ 759} (2016) 601,
  \href{http://dx.doi.org/10.1016/j.physletb.2016.06.023}{\doi{10.1016/j.physletb.2016.06.023}},
\href{http://www.arXiv.org/abs/1603.09222}{\texttt{arXiv:1603.09222}}.

\bibitem{CMS:2019raw}
\hrefCMSnoop {}{{CMS Collaboration}, ``{Measurements of differential Z boson
  production cross sections in proton-proton collisions at $ \sqrt{s} $ = 13
  TeV}'',} \textit{ JHEP} \textbf{ 12} (2019) 061,
  \href{http://dx.doi.org/10.1007/JHEP12(2019)061}{\doi{10.1007/JHEP12(2019)061}},
  \href{http://www.arXiv.org/abs/1909.04133}{\texttt{arXiv:1909.04133}}.

\bibitem{CMS:2020cph}
\hrefCMSnoop {}{{CMS Collaboration}, ``{Measurements of the W boson rapidity,
  helicity, double-differential cross sections, and charge asymmetry in
  $\Pp\Pp$ collisions at $\sqrt {s}$ = 13 TeV}'',} \textit{ Phys. Rev. D}
  \textbf{ 102} (2020) 092012,
  \href{http://dx.doi.org/10.1103/PhysRevD.102.092012}{\doi{10.1103/PhysRevD.102.092012}},
  \href{http://www.arXiv.org/abs/2008.04174}{\texttt{arXiv:2008.04174}}.

\bibitem{ATLAS:2024nrd}
\hrefCMSnoop {}{{ATLAS Collaboration}, ``{Precise measurements of W- and
  Z-boson transverse momentum spectra with the ATLAS detector using pp
  collisions at $\sqrt{s} = 5.02$ TeV and 13 TeV}'',} \textit{ Eur. Phys. J. C}
  \textbf{ 84} (2024) 1126,
  \href{http://dx.doi.org/10.1140/epjc/s10052-024-13414-0}{\doi{10.1140/epjc/s10052-024-13414-0}},
  \href{http://www.arXiv.org/abs/2404.06204}{\texttt{arXiv:2404.06204}}.

\bibitem{ATLAS:2024irg}
\hrefCMSnoop {}{{ATLAS Collaboration}, ``{Measurement of vector boson
  production cross sections and their ratios using pp collisions at s=13.6 TeV
  with the ATLAS detector}'',} \textit{ Phys. Lett. B} \textbf{ 854} (2024)
  138725,
  \href{http://dx.doi.org/10.1016/j.physletb.2024.138725}{\doi{10.1016/j.physletb.2024.138725}},
  \href{http://www.arXiv.org/abs/2403.12902}{\texttt{arXiv:2403.12902}}.

\bibitem{hepdata}
\hrefCMSnoop {}{}{HEPD}ata record for this analysis, 2024.
\newblock
  \href{http://dx.doi.org/10.17182/hepdata.153468}{\doi{10.17182/hepdata.153468}}.

\bibitem{CMS:2020cmk}
\hrefCMSnoop {}{{CMS Collaboration}, ``{Performance of the CMS Level-1 trigger
  in proton-proton collisions at $\sqrt{s} = 13$\,TeV}'',} \textit{ JINST}
  \textbf{ 15} (2020) P10017,
  \href{http://dx.doi.org/10.1088/1748-0221/15/10/P10017}{\doi{10.1088/1748-0221/15/10/P10017}},
  \href{http://www.arXiv.org/abs/2006.10165}{\texttt{arXiv:2006.10165}}.

\bibitem{Khachatryan:2016bia}
\hrefCMSnoop {}{{CMS Collaboration}, ``{The CMS trigger system}'',} \textit{
  JINST} \textbf{ 12} (2017) P01020,
  \href{http://dx.doi.org/10.1088/1748-0221/12/01/P01020}{\doi{10.1088/1748-0221/12/01/P01020}},
\href{http://www.arXiv.org/abs/1609.02366}{\texttt{arXiv:1609.02366}}.

\bibitem{Alwall:2014hca}
J.~Alwall\hrefCMSnoop {}{ { et~al.}, ``{The automated computation of tree-level
  and next-to-leading order differential cross sections, and their matching to
  parton shower simulations}'',} \textit{ JHEP} \textbf{ 07} (2014) 079,
  \href{http://dx.doi.org/10.1007/JHEP07(2014)079}{\doi{10.1007/JHEP07(2014)079}},
\href{http://www.arXiv.org/abs/1405.0301}{\texttt{arXiv:1405.0301}}.

\bibitem{Nason:2004rx}
\hrefCMSnoop {}{P.~Nason, ``A new method for combining {NLO QCD} with shower
  {Monte Carlo} algorithms'',} \textit{ JHEP} \textbf{ 11} (2004) 040,
  \href{http://dx.doi.org/10.1088/1126-6708/2004/11/040}{\doi{10.1088/1126-6708/2004/11/040}},
\href{http://www.arXiv.org/abs/hep-ph/0409146}{\texttt{arXiv:hep-ph/0409146}}.

\bibitem{Frixione:2007vw}
\hrefCMSnoop {}{S.~Frixione, P.~Nason, and C.~Oleari, ``{Matching NLO QCD
  computations with parton shower simulations: the POWHEG method}'',} \textit{
  JHEP} \textbf{ 11} (2007) 070,
  \href{http://dx.doi.org/10.1088/1126-6708/2007/11/070}{\doi{10.1088/1126-6708/2007/11/070}},
\href{http://www.arXiv.org/abs/0709.2092}{\texttt{arXiv:0709.2092}}.

\bibitem{Alioli:2008gx}
\hrefCMSnoop {}{S.~Alioli, P.~Nason, C.~Oleari, and E.~Re, ``{NLO} vector-boson
  production matched with shower in {POWHEG}'',} \textit{ JHEP} \textbf{ 07}
  (2008) 060,
  \href{http://dx.doi.org/10.1088/1126-6708/2008/07/060}{\doi{10.1088/1126-6708/2008/07/060}},
\href{http://www.arXiv.org/abs/0805.4802}{\texttt{arXiv:0805.4802}}.

\bibitem{Alioli:2010xd}
\hrefCMSnoop {}{S.~Alioli, P.~Nason, C.~Oleari, and E.~Re, ``{A general
  framework for implementing NLO calculations in shower Monte Carlo programs:
  the POWHEG BOX}'',} \textit{ JHEP} \textbf{ 06} (2010) 043,
  \href{http://dx.doi.org/10.1007/JHEP06(2010)043}{\doi{10.1007/JHEP06(2010)043}},
\href{http://www.arXiv.org/abs/1002.2581}{\texttt{arXiv:1002.2581}}.

\bibitem{Chiesa:2020ttl}
\hrefCMSnoop {}{M.~Chiesa, C.~Oleari, and E.~Re, ``{NLO QCD+NLO EW corrections
  to diboson production matched to parton shower}'',} \textit{ Eur. Phys. J. C}
  \textbf{ 80} (2020) 849,
  \href{http://dx.doi.org/10.1140/epjc/s10052-020-8419-3}{\doi{10.1140/epjc/s10052-020-8419-3}},
  \href{http://www.arXiv.org/abs/2005.12146}{\texttt{arXiv:2005.12146}}.

\bibitem{Frixione:2007nu}
\hrefCMSnoop {}{S.~Frixione, P.~Nason, and G.~Ridolfi, ``{The POWHEG-hvq manual
  version 1.0}'',} 2007.
  \href{http://www.arXiv.org/abs/0707.3081}{\texttt{arXiv:0707.3081}}.

\bibitem{Sjostrand:2014zea}
T.~Sj{\"o}strand\hrefCMSnoop {}{ { et~al.}, ``{An introduction to PYTHIA
  8.2}'',} \textit{ Comput. Phys. Commun.} \textbf{ 191} (2015) 159,
  \href{http://dx.doi.org/10.1016/j.cpc.2015.01.024}{\doi{10.1016/j.cpc.2015.01.024}},
\href{http://www.arXiv.org/abs/1410.3012}{\texttt{arXiv:1410.3012}}.

\bibitem{Skands:2014pea}
\hrefCMSnoop {}{P.~Skands, S.~Carrazza, and J.~Rojo, ``Tuning {PYTHIA} 8.1: the
  {Monash} 2013 tune'',} \textit{ Eur. Phys. J. C} \textbf{ 74} (2014) 3024,
  \href{http://dx.doi.org/10.1140/epjc/s10052-014-3024-y}{\doi{10.1140/epjc/s10052-014-3024-y}},
  \href{http://www.arXiv.org/abs/1404.5630}{\texttt{arXiv:1404.5630}}.

\bibitem{Sirunyan:2019dfx}
\hrefCMSnoop {}{{CMS Collaboration}, ``{Extraction and validation of a new set
  of CMS \PYTHIA~8 tunes from underlying-event measurements}'',} \textit{ Eur.
  Phys. J. C} \textbf{ 80} (2020) 4,
  \href{http://dx.doi.org/10.1140/epjc/s10052-019-7499-4}{\doi{10.1140/epjc/s10052-019-7499-4}},
\href{http://www.arXiv.org/abs/1903.12179}{\texttt{arXiv:1903.12179}}.

\bibitem{Agostinelli:2002hh}
\hrefCMSnoop {}{{GEANT4} Collaboration, ``{\GEANTfour}---a simulation
  toolkit'',} \textit{ Nucl. Instrum. Meth. A} \textbf{ 506} (2003) 250,
  \href{http://dx.doi.org/10.1016/S0168-9002(03)01368-8}{\doi{10.1016/S0168-9002(03)01368-8}}.

\bibitem{Sirunyan:2017ulk}
\hrefCMSnoop {}{{CMS Collaboration}, ``{Particle-flow reconstruction and global
  event description with the CMS detector}'',} \textit{ JINST} \textbf{ 12}
  (2017) P10003,
  \href{http://dx.doi.org/10.1088/1748-0221/12/10/P10003}{\doi{10.1088/1748-0221/12/10/P10003}},
\href{http://www.arXiv.org/abs/1706.04965}{\texttt{arXiv:1706.04965}}.

\bibitem{CMS-TDR-15-02}
\href {http://cds.cern.ch/record/2020886}{{CMS Collaboration}, ``Technical
  proposal for the {Phase-II} upgrade of the {Compact Muon Solenoid}'',} CMS
  Technical Proposal CERN-LHCC-2015-010, CMS-TDR-15-02, 2015.

\bibitem{Cacciari:2008gp}
\hrefCMSnoop {}{M.~Cacciari, G.~P. Salam, and G.~Soyez, ``{The anti-\kt jet
  clustering algorithm}'',} \textit{ JHEP} \textbf{ 04} (2008) 063,
  \href{http://dx.doi.org/10.1088/1126-6708/2008/04/063}{\doi{10.1088/1126-6708/2008/04/063}},
  \href{http://www.arXiv.org/abs/0802.1189}{\texttt{arXiv:0802.1189}}.

\bibitem{Cacciari:2011ma}
\hrefCMSnoop {}{M.~Cacciari, G.~P. Salam, and G.~Soyez, ``{FastJet user
  manual}'',} \textit{ Eur. Phys. J. C} \textbf{ 72} (2012) 1896,
  \href{http://dx.doi.org/10.1140/epjc/s10052-012-1896-2}{\doi{10.1140/epjc/s10052-012-1896-2}},
\href{http://www.arXiv.org/abs/1111.6097}{\texttt{arXiv:1111.6097}}.

\bibitem{CMS:2019ctu}
\hrefCMSnoop {}{{CMS Collaboration}, ``Performance of missing transverse
  momentum reconstruction in proton-proton collisions at $\sqrt{s} = 13$\,{TeV}
  using the {CMS} detector'',} \textit{ JINST} \textbf{ 14} (2019) P07004,
  \href{http://dx.doi.org/10.1088/1748-0221/14/07/P07004}{\doi{10.1088/1748-0221/14/07/P07004}},
\href{http://www.arXiv.org/abs/1903.06078}{\texttt{arXiv:1903.06078}}.

\bibitem{CMS:2020uim}
\hrefCMSnoop {}{{CMS Collaboration}, ``{Electron and photon reconstruction and
  identification with the CMS experiment at the CERN LHC}'',} \textit{ JINST}
  \textbf{ 16} (2021) P05014,
  \href{http://dx.doi.org/10.1088/1748-0221/16/05/P05014}{\doi{10.1088/1748-0221/16/05/P05014}},
  \href{http://www.arXiv.org/abs/2012.06888}{\texttt{arXiv:2012.06888}}.

\bibitem{CMS-DP-2020-021}
\href {https://cds.cern.ch/record/2717925}{{CMS Collaboration}, ``{ECAL} 2016
  refined calibration and {Run 2} summary plots'',} CMS Detector Performance
  Note CMS-DP-2020-021, 2020.

\bibitem{Sirunyan:2018fpa}
\hrefCMSnoop {}{{CMS Collaboration}, ``{Performance of the CMS muon detector
  and muon reconstruction with proton-proton collisions at $\sqrt{s}=$ 13
  TeV}'',} \textit{ JINST} \textbf{ 13} (2018) P06015,
  \href{http://dx.doi.org/10.1088/1748-0221/13/06/P06015}{\doi{10.1088/1748-0221/13/06/P06015}},
\href{http://www.arXiv.org/abs/1804.04528}{\texttt{arXiv:1804.04528}}.

\bibitem{CMS:2020ebo}
\hrefCMSnoop {}{{CMS Collaboration}, ``{Pileup mitigation at CMS in 13\TeV
  data}'',} \textit{ JINST} \textbf{ 15} (2020) P09018,
  \href{http://dx.doi.org/10.1088/1748-0221/15/09/P09018}{\doi{10.1088/1748-0221/15/09/P09018}},
  \href{http://www.arXiv.org/abs/2003.00503}{\texttt{arXiv:2003.00503}}.

\bibitem{Bodek:2012id}
A.~Bodek\hrefCMSnoop {}{ { et~al.}, ``Extracting muon momentum scale
  corrections for hadron collider experiments'',} \textit{ Eur. Phys. J. C}
  \textbf{ 72} (2012) 2194,
  \href{http://dx.doi.org/10.1140/epjc/s10052-012-2194-8}{\doi{10.1140/epjc/s10052-012-2194-8}},
\href{http://www.arXiv.org/abs/1208.3710}{\texttt{arXiv:1208.3710}}.

\bibitem{CMS:2024ppo}
\hrefCMSnoop {}{{CMS Collaboration}, ``{Performance of the CMS electromagnetic
  calorimeter in pp collisions at $\sqrt{s}$ = 13 TeV}'',} 2024.
  \href{http://www.arXiv.org/abs/2403.15518}{\texttt{arXiv:2403.15518}}.

\bibitem{Hou:2019efy}
\hrefCMSnoop {}{T.-J. Hou { et~al.}, ``{New CTEQ global analysis of quantum
  chromodynamics with high-precision data from the LHC}'',} \textit{ Phys. Rev.
  D} \textbf{ 103} (2021) 014013,
  \href{http://dx.doi.org/10.1103/PhysRevD.103.014013}{\doi{10.1103/PhysRevD.103.014013}},
  \href{http://www.arXiv.org/abs/1912.10053}{\texttt{arXiv:1912.10053}}.

\bibitem{CAT-23-001}
\hrefCMSnoop {}{{CMS Collaboration}, ``The {CMS} statistical analysis and
  combination tool: {\textsc{Combine}}'',} 2024.
  \href{http://www.arXiv.org/abs/2404.06614}{\texttt{arXiv:2404.06614}}.
  Submitted to \textit{Comput. Softw. Big Sci.}

\bibitem{CMS-PAS-LUM-17-004}
\href {https://cds.cern.ch/record/2621960}{{CMS Collaboration}, ``{CMS
  luminosity measurement for the 2017 data-taking period at $\sqrt{s} =
  13~\mathrm{TeV}$}'',} CMS Physics Analysis Summary CMS-PAS-LUM-17-004, 2017.

\bibitem{CMS-PAS-LUM-19-001}
\href {http://cds.cern.ch/record/2765655/}{{CMS Collaboration}, ``{Luminosity
  measurement in proton-proton collisions at 5.02 TeV in 2017 at CMS}'',} CMS
  Physics Analysis Summary CMS-PAS-LUM-19-001, 2021.

\bibitem{Golonka:2005pn}
\hrefCMSnoop {}{P.~Golonka and Z.~Was, ``{PHOTOS Monte Carlo: A precision tool
  for QED corrections in $\PZ$ and $\PW$ decays}'',} \textit{ Eur. Phys. J. C}
  \textbf{ 45} (2006) 97,
  \href{http://dx.doi.org/10.1140/epjc/s2005-02396-4}{\doi{10.1140/epjc/s2005-02396-4}},
\href{http://www.arXiv.org/abs/hep-ph/0506026}{\texttt{arXiv:hep-ph/0506026}}.

\bibitem{BARLOW1993219}
\hrefCMSnoop {}{R.~Barlow and C.~Beeston, ``Fitting using finite {Monte Carlo}
  samples'',} \textit{ Comput. Phys. Comm.} \textbf{ 77} (1993) 219,
  \href{http://dx.doi.org/10.1016/0010-4655(93)90005-W}{\doi{10.1016/0010-4655(93)90005-W}}.

\bibitem{Collins:1984kg}
\hrefCMSnoop {}{J.~C. Collins, D.~E. Soper, and G.~Sterman, ``{Transverse
  momentum distribution in Drell-Yan pair and W and Z boson production}'',}
  \textit{ Nucl. Phys. B} \textbf{ 250} (1985) 199,
\href{http://dx.doi.org/10.1016/0550-3213(85)90479-1}{\doi{10.1016/0550-3213(85)90479-1}}.

\bibitem{Gleisberg:2008ta}
T.~Gleisberg\hrefCMSnoop {}{ { et~al.}, ``{Event generation with SHERPA
  1.1}'',} \textit{ JHEP} \textbf{ 02} (2009) 007,
  \href{http://dx.doi.org/10.1088/1126-6708/2009/02/007}{\doi{10.1088/1126-6708/2009/02/007}},
\href{http://www.arXiv.org/abs/0811.4622}{\texttt{arXiv:0811.4622}}.

\bibitem{Bahr:2008pv}
M.~B{\"a}hr\hrefCMSnoop {}{ { et~al.}, ``Herwig++ physics and manual'',}
  \textit{ Eur. Phys. J. C} \textbf{ 58} (2008) 639,
  \href{http://dx.doi.org/10.1140/epjc/s10052-008-0798-9}{\doi{10.1140/epjc/s10052-008-0798-9}},
\href{http://www.arXiv.org/abs/0803.0883}{\texttt{arXiv:0803.0883}}.

\bibitem{Frixione:2002ik}
\hrefCMSnoop {}{S.~Frixione and B.~R. Webber, ``{Matching NLO QCD computations
  and parton shower simulations}'',} \textit{ JHEP} \textbf{ 06} (2002) 029,
  \href{http://dx.doi.org/10.1088/1126-6708/2002/06/029}{\doi{10.1088/1126-6708/2002/06/029}},
\href{http://www.arXiv.org/abs/hep-ph/0204244}{\texttt{arXiv:hep-ph/0204244}}.

\bibitem{Balazs:1995nz}
\hrefCMSnoop {}{C.~Bal{\'a}zs, J.-W. Qiu, and C.~P. Yuan, ``{Effects of QCD
  resummation on distributions of leptons from the decay of electroweak vector
  bosons}'',} \textit{ Phys. Lett. B} \textbf{ 355} (1995) 548,
  \href{http://dx.doi.org/10.1016/0370-2693(95)00726-2}{\doi{10.1016/0370-2693(95)00726-2}},
\href{http://www.arXiv.org/abs/hep-ph/9505203}{\texttt{arXiv:hep-ph/9505203}}.

\bibitem{Catani:2015vma}
\hrefCMSnoop {}{S.~Catani, D.~de~Florian, G.~Ferrera, and M.~Grazzini,
  ``{Vector boson production at hadron colliders: transverse-momentum
  resummation and leptonic decay}'',} \textit{ JHEP} \textbf{ 12} (2015) 047,
  \href{http://dx.doi.org/10.1007/JHEP12(2015)047}{\doi{10.1007/JHEP12(2015)047}},
\href{http://www.arXiv.org/abs/1507.06937}{\texttt{arXiv:1507.06937}}.

\bibitem{Camarda:2019zyx}
\hrefCMSnoop {}{S.~Camarda { et~al.}, ``{DYTurbo: Fast predictions for
  Drell-Yan processes}'',} \textit{ Eur. Phys. J. C} \textbf{ 80} (2020) 251,
  \href{http://dx.doi.org/10.1140/epjc/s10052-020-7757-5}{\doi{10.1140/epjc/s10052-020-7757-5}},
  \href{http://www.arXiv.org/abs/1910.07049}{\texttt{arXiv:1910.07049}}.
  [Erratum: \DOI{10.1140/epjc/s10052-020-7972-0}].

\bibitem{Camarda:2021ict}
\hrefCMSnoop {}{S.~Camarda, L.~Cieri, and G.~Ferrera, ``{Drell\textendash{}Yan
  lepton-pair production: $\mathrm{q}_\mathrm{T}$ resummation at N$^3$LL
  accuracy and fiducial cross sections at N$^3$LO}'',} \textit{ Phys. Rev. D}
  \textbf{ 104} (2021) L111503,
  \href{http://dx.doi.org/10.1103/PhysRevD.104.L111503}{\doi{10.1103/PhysRevD.104.L111503}},
  \href{http://www.arXiv.org/abs/2103.04974}{\texttt{arXiv:2103.04974}}.

\bibitem{Camarda:2021jsw}
\hrefCMSnoop {}{S.~Camarda, L.~Cieri, and G.~Ferrera, ``{Fiducial perturbative
  power corrections within the $\mathrm{q}_\mathrm{T}$ subtraction
  formalism}'',} \textit{ Eur. Phys. J. C} \textbf{ 82} (2022) 575,
  \href{http://dx.doi.org/10.1140/epjc/s10052-022-10510-x}{\doi{10.1140/epjc/s10052-022-10510-x}},
  \href{http://www.arXiv.org/abs/2111.14509}{\texttt{arXiv:2111.14509}}.

\bibitem{Camarda:2023dqn}
\hrefCMSnoop {}{S.~Camarda, L.~Cieri, and G.~Ferrera, ``{Drell-Yan lepton-pair
  production: $\mathrm{q}_\mathrm{T}$ resummation at N$^4$LL accuracy}'',}
  \textit{ Phys. Lett. B} \textbf{ 845} (2023) 138125,
  \href{http://dx.doi.org/10.1016/j.physletb.2023.138125}{\doi{10.1016/j.physletb.2023.138125}},
  \href{http://www.arXiv.org/abs/2303.12781}{\texttt{arXiv:2303.12781}}.

\bibitem{NNPDF:2021njg}
\hrefCMSnoop {}{{NNPDF} Collaboration, ``{The path to proton structure at 1\%
  accuracy}'',} \textit{ Eur. Phys. J. C} \textbf{ 82} (2022) 428,
  \href{http://dx.doi.org/10.1140/epjc/s10052-022-10328-7}{\doi{10.1140/epjc/s10052-022-10328-7}},
  \href{http://www.arXiv.org/abs/2109.02653}{\texttt{arXiv:2109.02653}}.

\bibitem{Lepage:1977sw}
\hrefCMSnoop {}{G.~P. Lepage, ``{A New Algorithm for Adaptive Multidimensional
  Integration}'',} \textit{ J. Comput. Phys.} \textbf{ 27} (1978) 192,
  \href{http://dx.doi.org/10.1016/0021-9991(78)90004-9}{\doi{10.1016/0021-9991(78)90004-9}}.

\bibitem{UA1:1987dfv}
\hrefCMSnoop {}{{UA1} Collaboration, ``{Intermediate vector boson cross
  sections at the CERN super proton synchrotron collider and the number of
  neutrino types}'',} \textit{ Phys. Lett. B} \textbf{ 198} (1987) 271,
  \href{http://dx.doi.org/10.1016/0370-2693(87)91510-3}{\doi{10.1016/0370-2693(87)91510-3}}.

\bibitem{UA2:1990mxf}
\hrefCMSnoop {}{{UA2} Collaboration, ``{Measurement of W and Z production cross
  sections at the CERN $\mathrm{\bar{p}p}$ collider}'',} \textit{ Z. Phys. C}
  \textbf{ 47} (1990) 11,
  \href{http://dx.doi.org/10.1007/BF01551906}{\doi{10.1007/BF01551906}}.

\bibitem{D0:1999qdf}
\hrefCMSnoop {}{{D0} Collaboration, ``{Extraction of the width of the W boson
  from measurements of $\sigma(p\bar{p} \to W + X) \times B(W \to e \nu)$ and
  $\sigma(p\bar{p} \to Z + X) \times B(Z \to e e)$ and their ratio}'',}
  \textit{ Phys. Rev. D} \textbf{ 61} (2000) 072001,
  \href{http://dx.doi.org/10.1103/PhysRevD.61.072001}{\doi{10.1103/PhysRevD.61.072001}},
  \href{http://www.arXiv.org/abs/hep-ex/9906025}{\texttt{arXiv:hep-ex/9906025}}.

\bibitem{CDF:2004sns}
\hrefCMSnoop {}{{CDF} Collaboration, ``{First measurements of inclusive $W$ and
  $Z$ cross sections from Run II of the Tevatron collider}'',} \textit{ Phys.
  Rev. Lett.} \textbf{ 94} (2005) 091803,
  \href{http://dx.doi.org/10.1103/PhysRevLett.94.091803}{\doi{10.1103/PhysRevLett.94.091803}},
  \href{http://www.arXiv.org/abs/hep-ex/0406078}{\texttt{arXiv:hep-ex/0406078}}.

\end{thebibliography}\endgroup
